\documentclass[doublecol]{epl2} 
\usepackage{graphicx}
\usepackage{subfigure}

\newcommand{\ER}{Erd{\H o}s-R{\'e}nyi }
\newcommand{\BA}{Barab\'{a}si-Albert }
\newcommand{\MR}{Molloy-Reed }

\begin{document}

\title{Portraits of complex networks}

\author{James P.~Bagrow\inst{1} \and Erik M.~Bollt\inst{2,1} \and Joseph D.~Skufca\inst{2} \and Daniel ben-Avraham\inst{1}}
\shortauthor{J.~Bagrow \etal}

\institute{
	\inst{1} Department of Physics, Clarkson University, Potsdam NY 13699-5820 USA\\
	\inst{2} Department of Math and Computer Science, Clarkson University, Potsdam, NY 13699-5815 USA
}

\date{January 27, 2008}

\abstract{We propose a method for characterizing large complex networks by introducing a new matrix structure, unique for a given network, which encodes structural information; provides useful visualization, even for very large networks; and allows for rigorous statistical comparison between networks.  Dynamic processes such as percolation can be visualized using animation.}


\maketitle

\section{Introduction}
Large, complex stochastic networks are conspicuous in science and everyday life and have attracted a great deal of interest~\cite{wattsStrogatz:Nature:SmallWorld,movieActors,BA:nature:metabolic}.
A difficult problem when studying networks is that of comparison and identification.  Given two networks, how similar are they?  Could they have arisen from the same generating mechanism?  Given a real-world network, such as a protein-protein interaction network, or an electric power grid, say, how can one determine which stochastic network model most accurately captures its relevant structure?  Is there
a reasonable way to illustrate what a particular network looks like?

\begin{figure} 
\centering
	\subfigure[][]{
	    \includegraphics[width=0.20\textwidth]{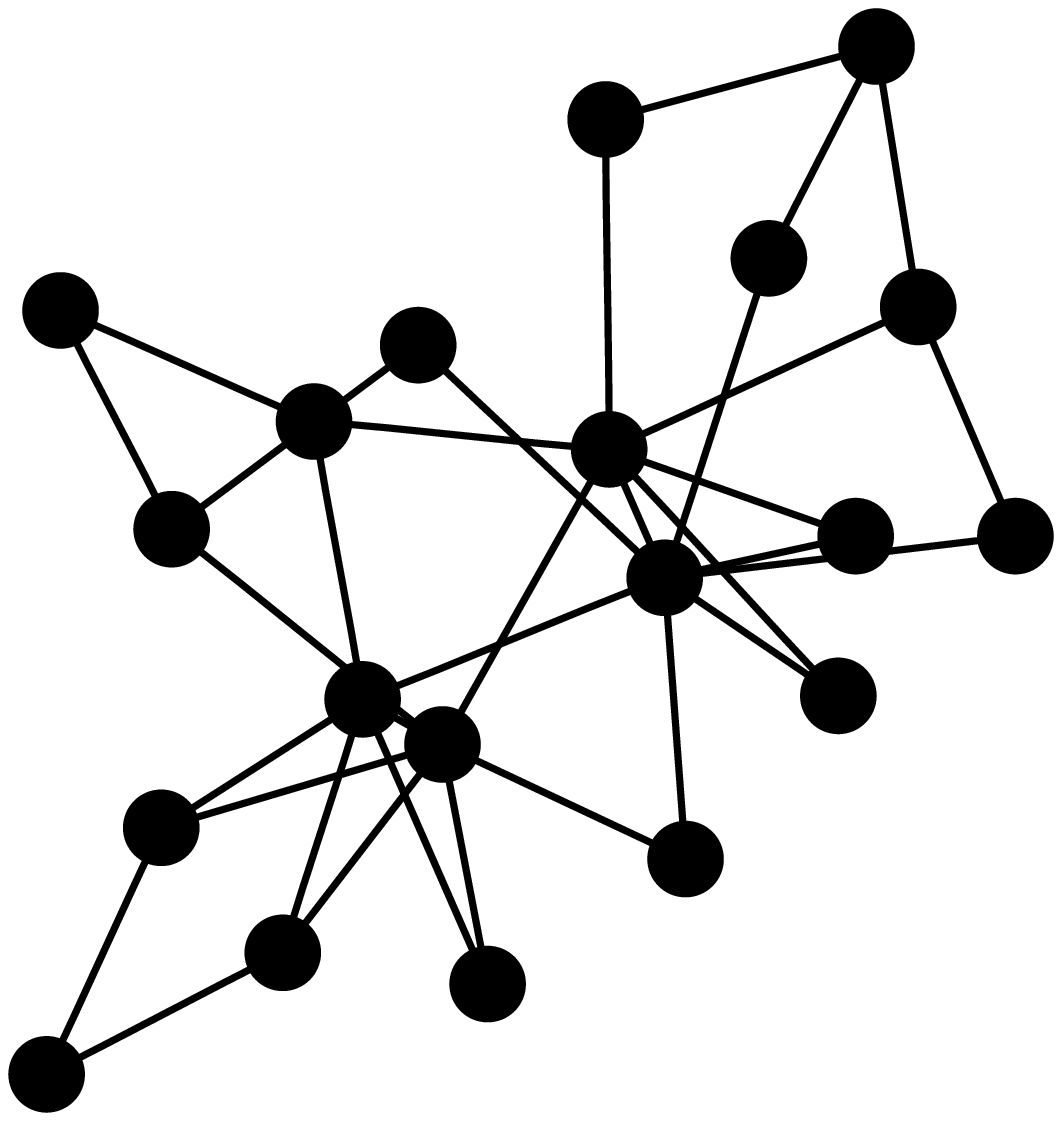}
	}
	\subfigure[][]{
	    \includegraphics[width=0.20\textwidth,]{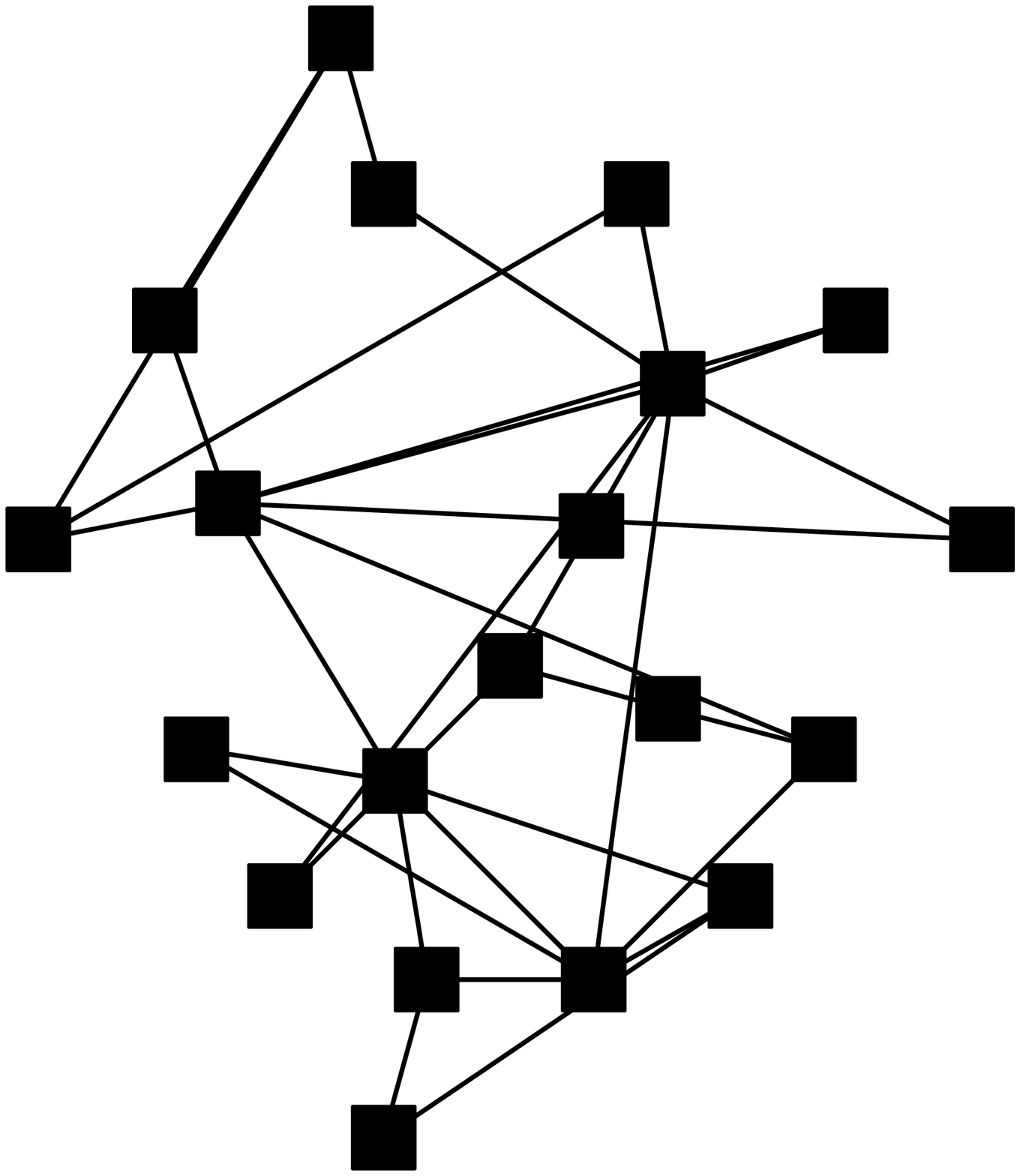}
	}
	\subfigure[][]{
	    \includegraphics[width=0.20\textwidth,trim=0 0 0 0,clip,]{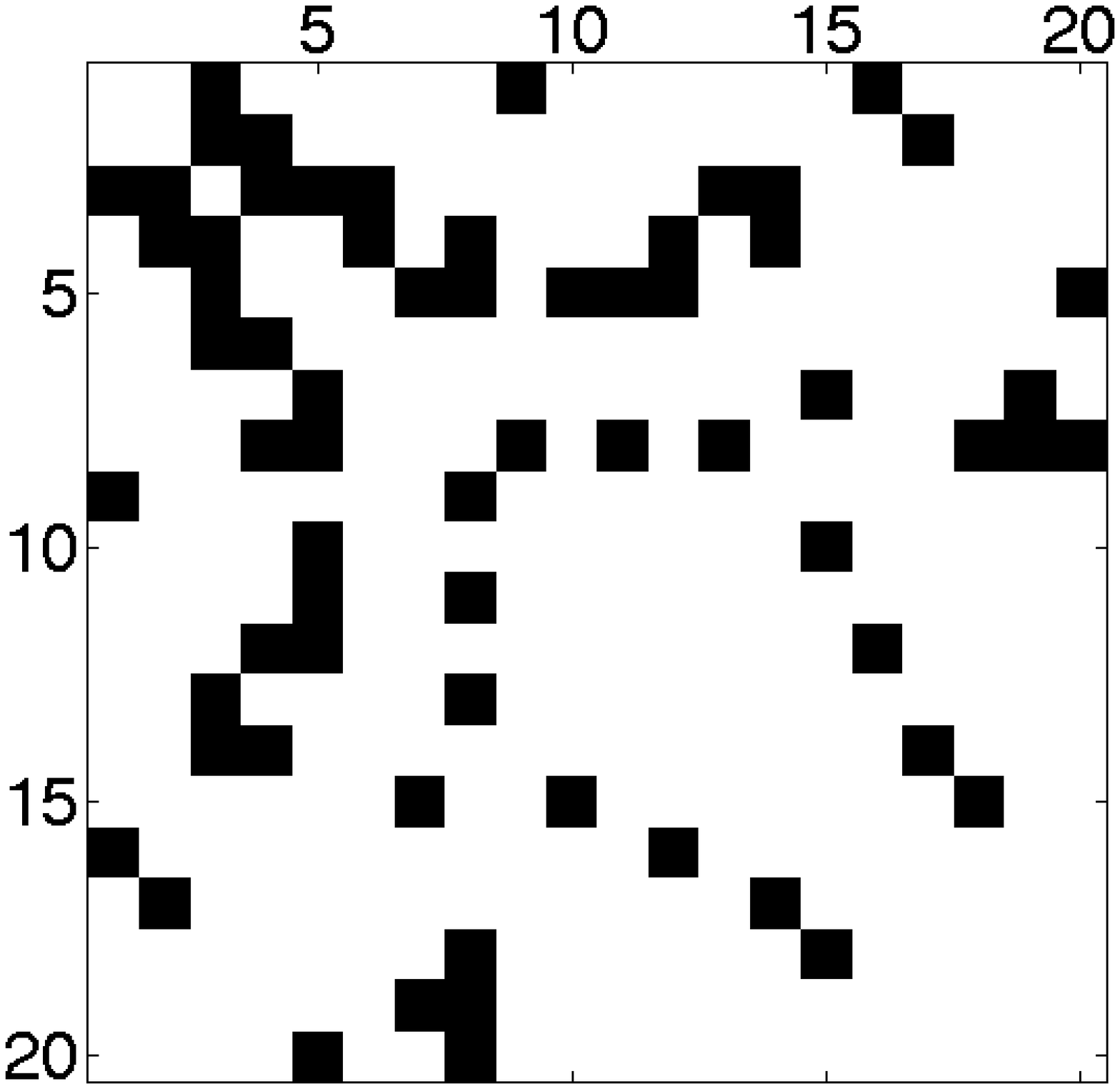}
	}
	\subfigure[][]{
	    \includegraphics[width=0.20\textwidth,trim=0 0 0 0,clip,]{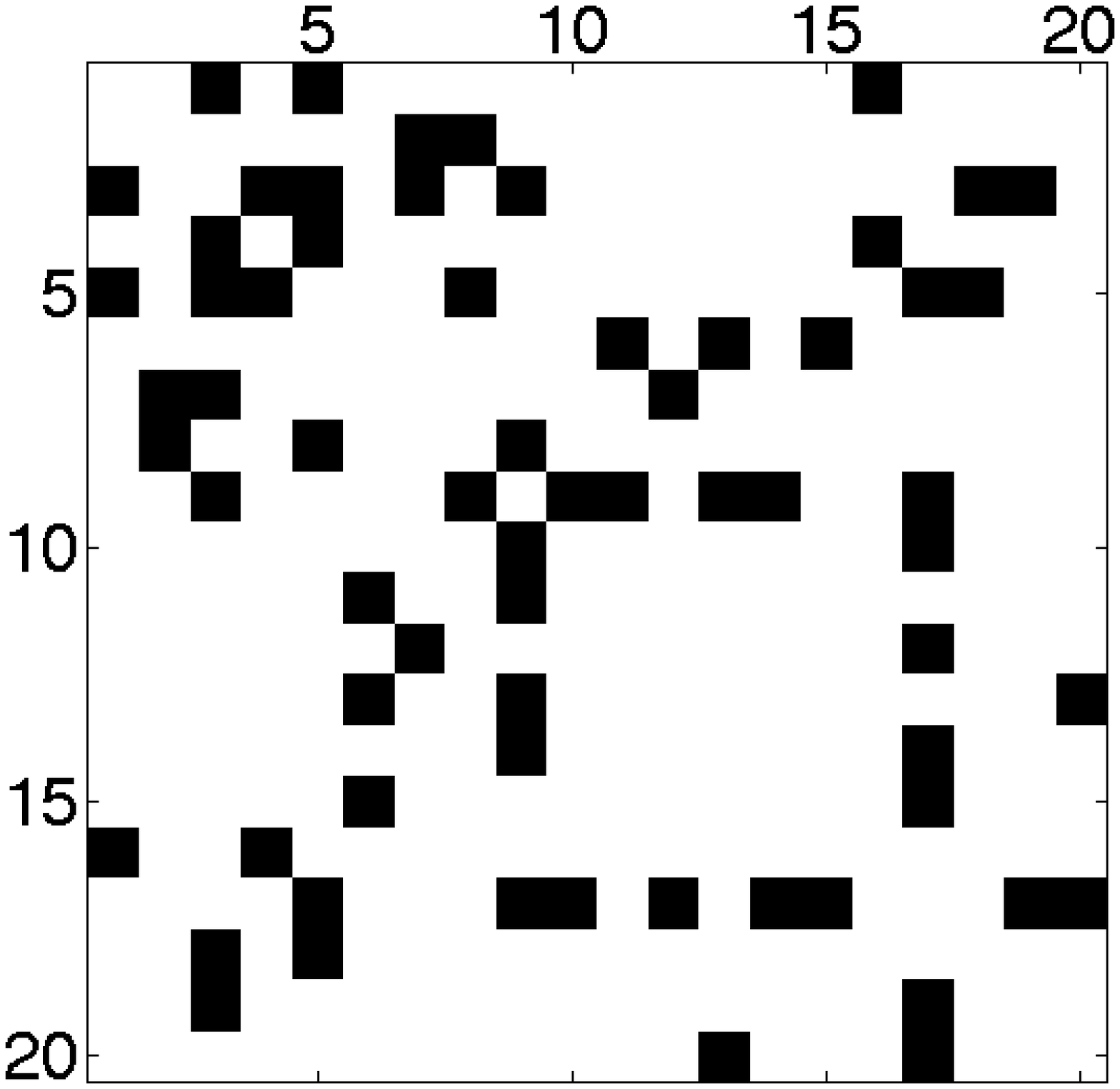}}
	\caption[]{\label{fig:ProblemsRepresenting}Planar embeddings and adjacency matrices for a small network.  It is difficult to tell visually that these represent the same network, even at such a small size. }
\end{figure}

A network, or graph, is characterized completely by its  {\it adjacency matrix\/} --- an $N\times N$ matrix whose nonzero entries denote the various links between the graph's $N$ nodes.  This representation, however, is not unique, in that it depends on the actual labeling of the nodes, and graph {\it isomorphs\/}  (identical graphs with permuted labels) cannot  be readily distinguished from one another~\cite{book:GI}.  The same is true of graphical representations, where node placement is arbitrary (Fig.\ \ref{fig:ProblemsRepresenting}).

In this letter, we propose a new method for recognizing and characterizing large complex networks that is independent of labeling  and circumvents the problem of graph isomorphism.  For each network we compute its $B$-matrix: a signature that represents the network reliably and serves as its `portrait.'   We thus have a means for recognizing networks at a glance and judge their differences and similarities, for the first time,  enormously increasing our understanding and intuition~\cite{remark1}.  
We also introduce a ``distance,'' derived from the $B$-matrix, that quantifies network differences, rendering comparisons mathematically meaningful.  One important application is to the comparison of phylogenetic  trees representing various organisms \cite{phylo}.

\section{Portraits}
A graph $G$ consists of a finite set of nodes, or vertices, $V=\{v_1,v_2,...,v_N\}$,  and a set of edges, or links, between pairs of vertices, $E=\{(v_i,v_j)\}$.  In applications, the vertices label elements of a network, and edges denote relationships between elements.  
The number of links, $k_i$, connected to a vertex $v_i$ is the {\it degree\/} of the vertex.  Much recent interest has focused on  {\it scale-free\/} networks, which exhibit  a power-law degree distribution, $P(k)\sim k^{-\gamma}$.  Despite its strong influence on various properties, the degree distribution is but one of many characteristics.  Two large networks may possess
similar degree distributions yet differ widely in {\it clustering\/} (the extent to which neighbors of a node connect to one another)~\cite{wattsStrogatz:Nature:SmallWorld}, {\it assortativity\/} (the frequency of connections between nodes of like degrees)~\cite{newmanAssortativity}, and other important properties.

\begin{figure}[] 
		\subfigure[][]{
		\label{subfig:actorLinear}
\includegraphics[trim=8 4 8 4,clip,width=0.46\textwidth]{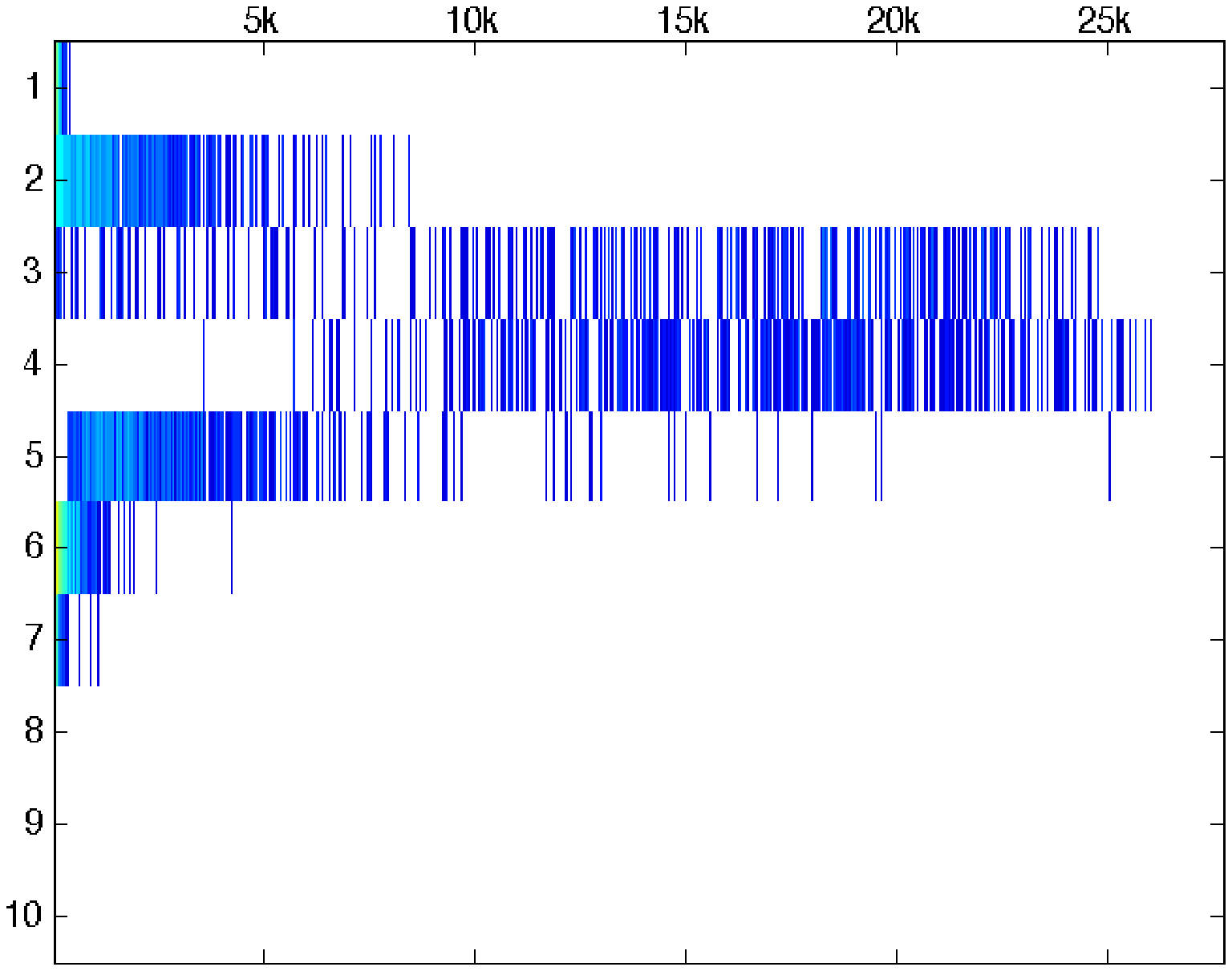}
	}
	\subfigure[][]{
	\label{subfig:actorLog}
\includegraphics[trim=8 4 8 4,clip,width=0.46\textwidth]{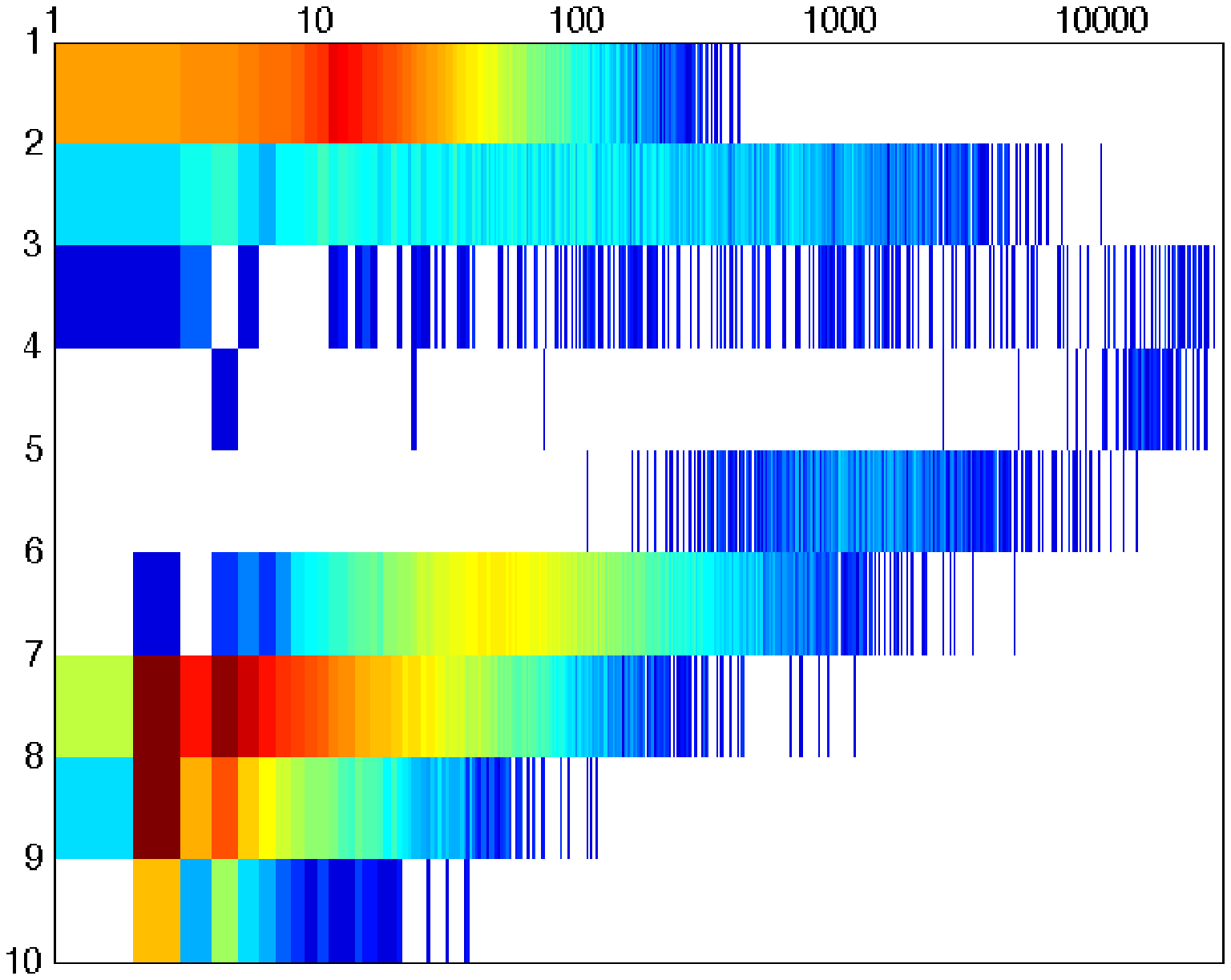}
	}
	\caption{\label{fig:introduceBmatrix}(color online) \subref{subfig:actorLinear} A $B$-Matrix with a logarithmic color scale (the white background indicates zero elements of $B$).  The degree distribution is slightly visible in the first row.  The ``turning point'' about row 4 represents finite-size effects.  Shown is the network of the $10\%$ most connected actors on IMDB~\cite{movieActors}. \subref{subfig:actorLog} The same matrix with a logarithmic horizontal axis.  The degree distribution is now clearly visible.}
\end{figure}

\begin{figure} 
\centering
	\subfigure[][]{
		\label{subfig:singleInstanceER}
	    \includegraphics[trim=8 4 8 4,clip,width=0.22\textwidth]{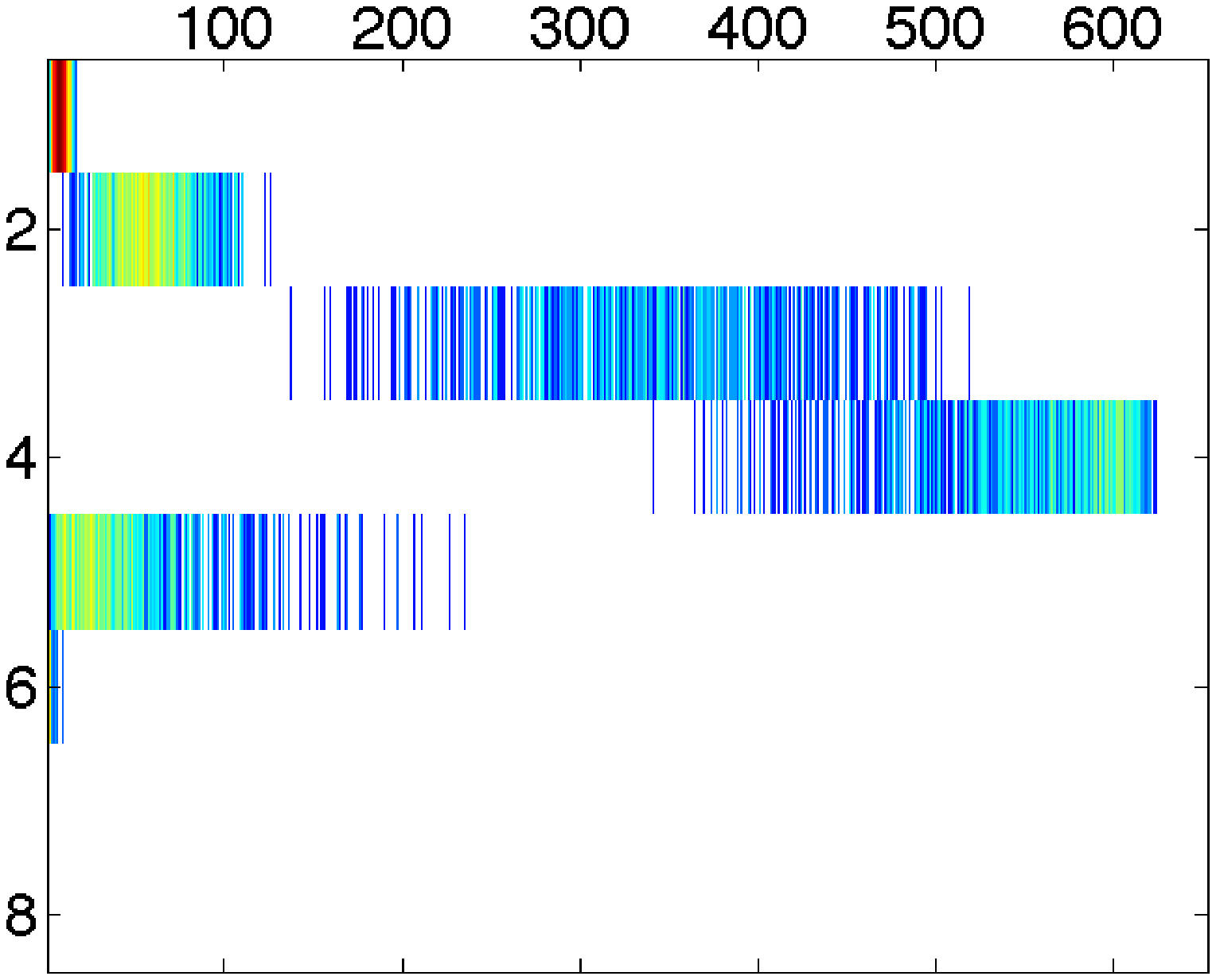}
	}
	\subfigure[][]{
	    \includegraphics[trim=8 4 8 4,clip,width=0.22\textwidth]{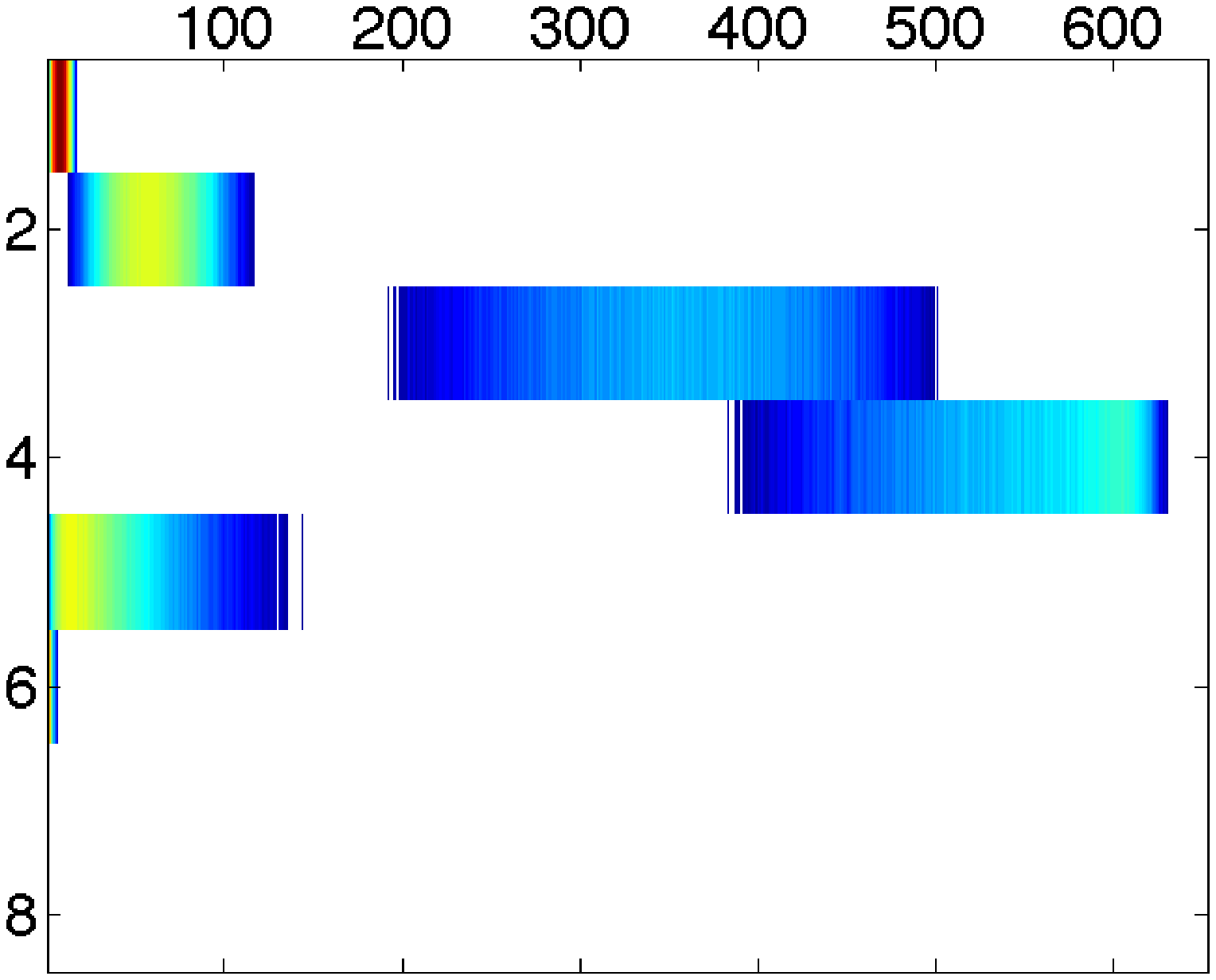}\label{subfig:averageER}
	}
	\subfigure[][]{
		\label{subfig:ERbelowPerc}
	    \includegraphics[trim=8 4 8 4,clip,width=0.22\textwidth]{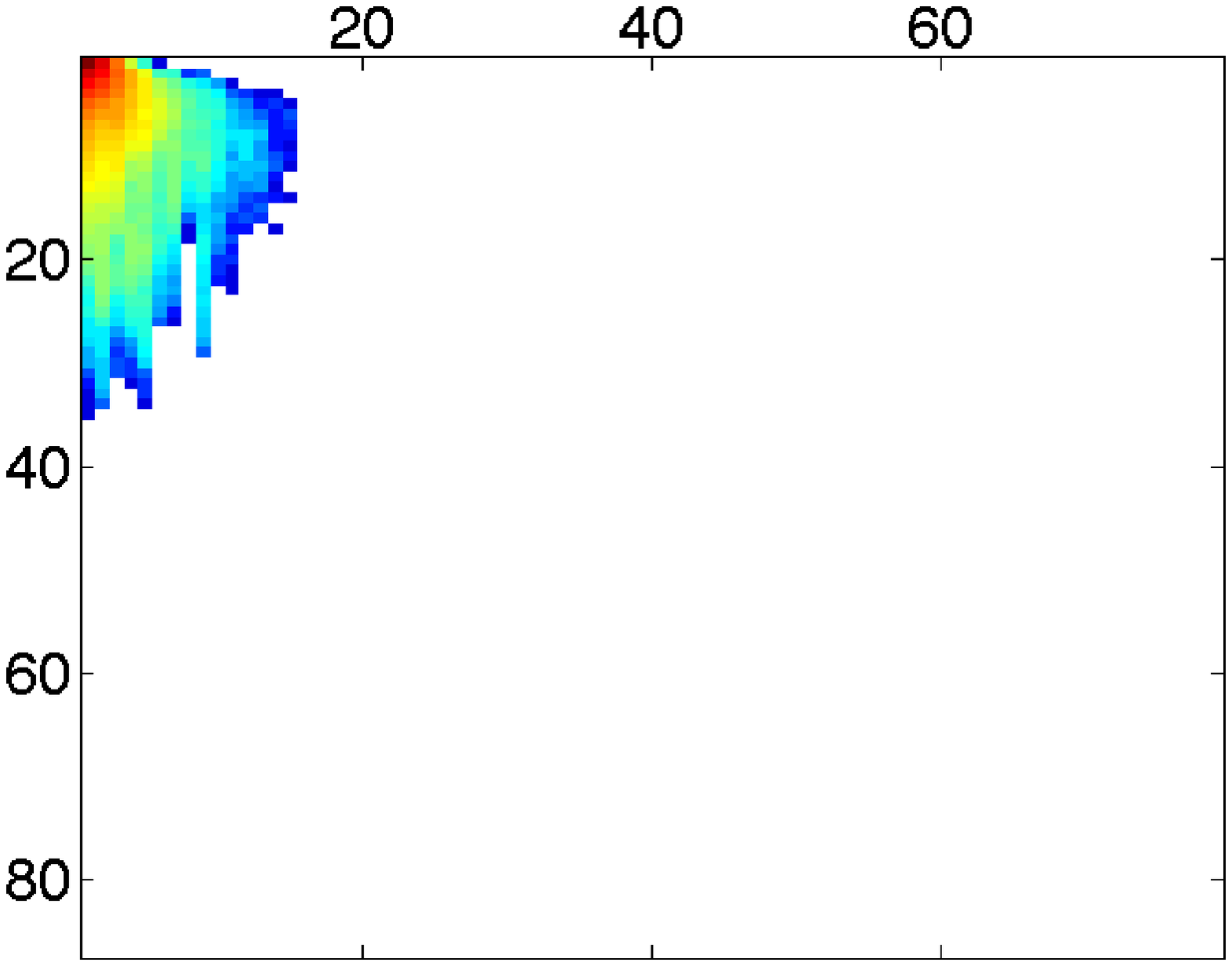}
	}
	\subfigure[][]{
	\label{subfig:ERatPerc}
	    \includegraphics[trim=8 4 8 4,clip,width=0.22\textwidth]{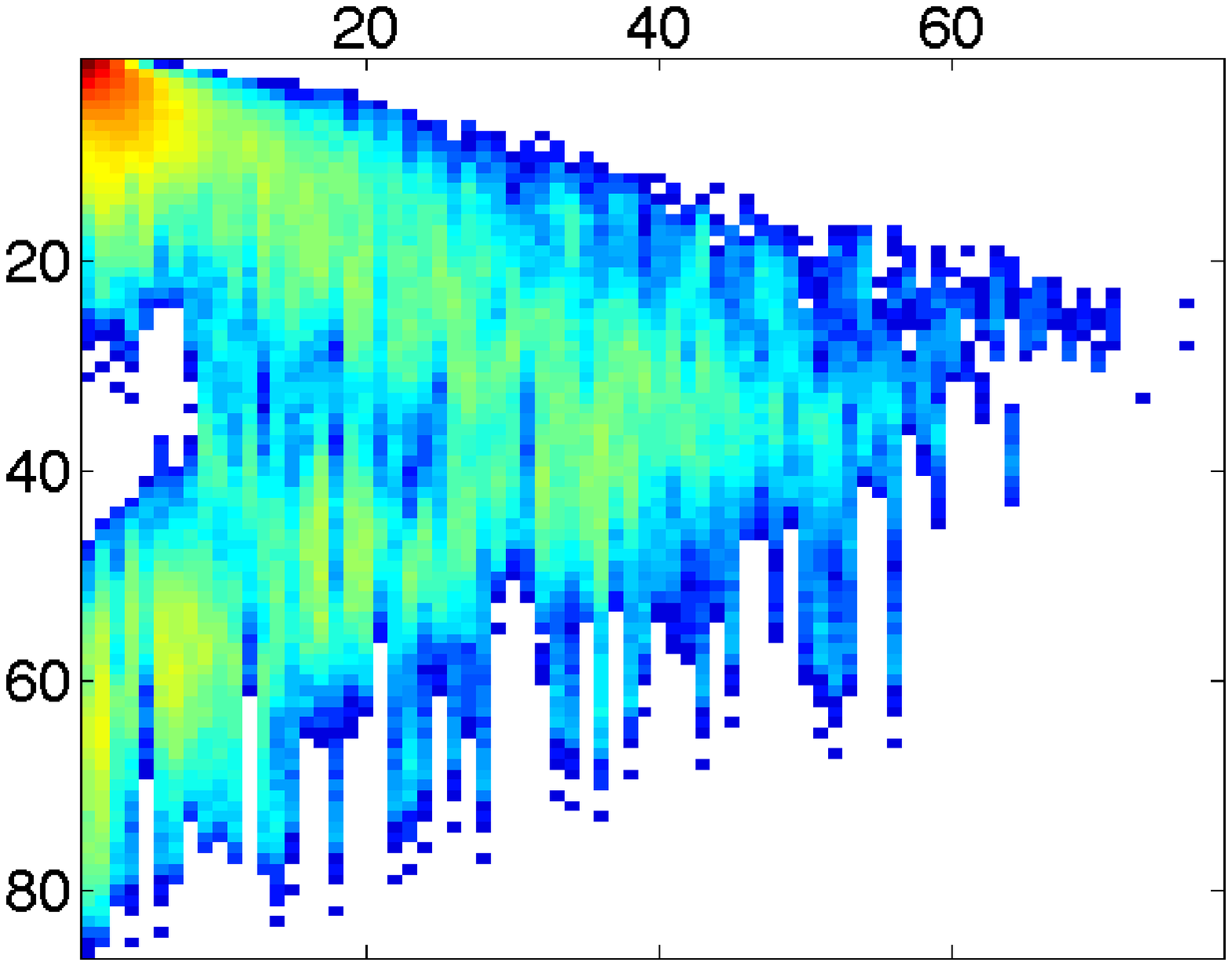}
	}
	\caption[]{\label{fig:variousERgraphs}(color online) Erd{\H o}s-R{\'e}nyi (ER) graphs~\protect{\cite{ERgraphs}}.  \subref{subfig:singleInstanceER} One graph with $N=1000$ nodes and $p=0.008$. \subref{subfig:averageER} The average of 100 graphs from \subref{subfig:singleInstanceER}.  Visualizing percolation: $N=10^4$ \subref{subfig:ERbelowPerc} below percolation, $p=(1.1N)^{-1}$; \subref{subfig:ERatPerc} at percolation, $p=1/N$.  }
\end{figure}

\begin{figure} 
\centering
	\subfigure[][]{
	\label{subfig:NWS_1over12}
	    \includegraphics[trim=8 4 8 4,clip,width=0.22\textwidth]{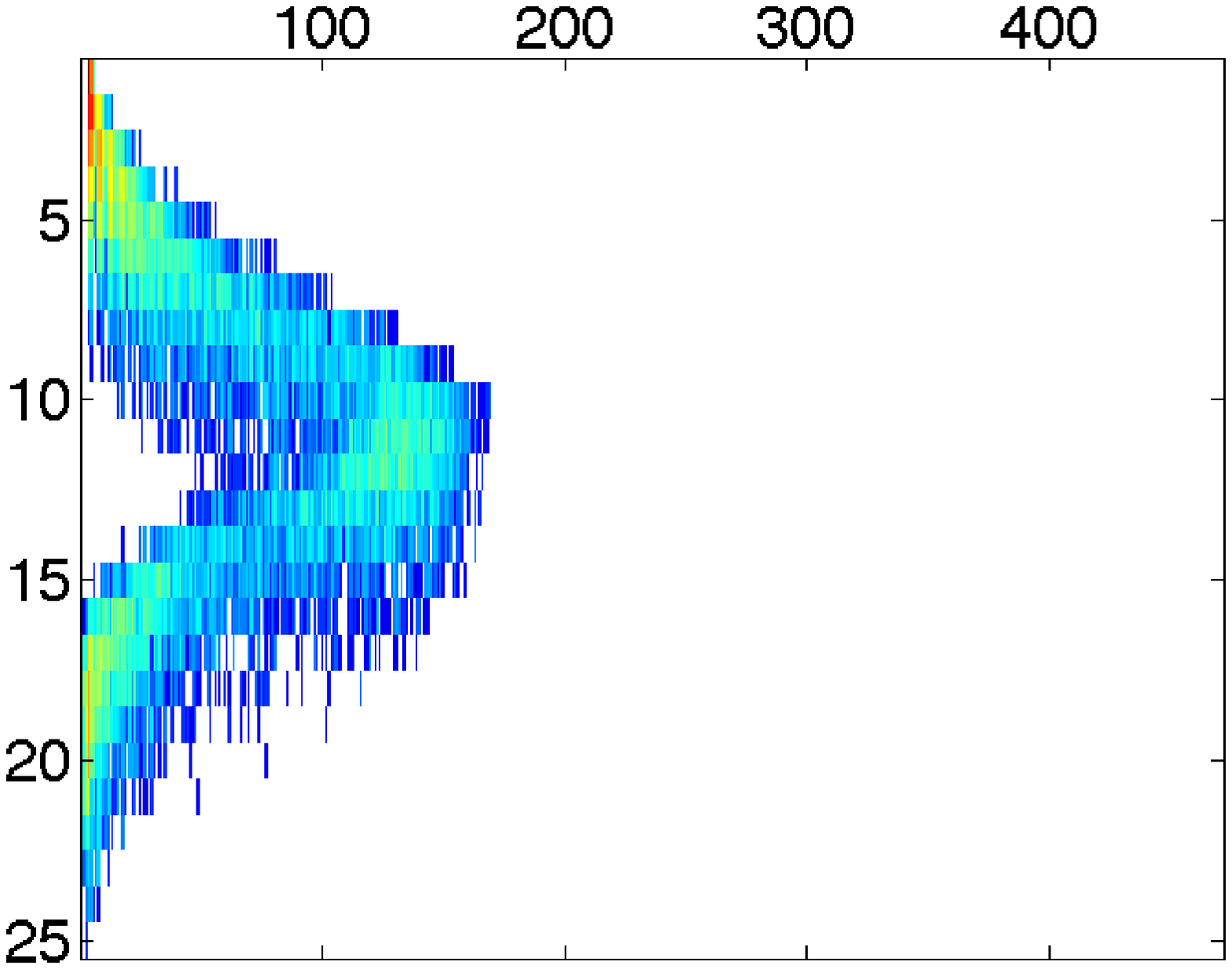}
	}
	\subfigure[][]{
	\label{subfig:NWS_1over9}
	    \includegraphics[trim=8 4 8 4,clip,width=0.22\textwidth]{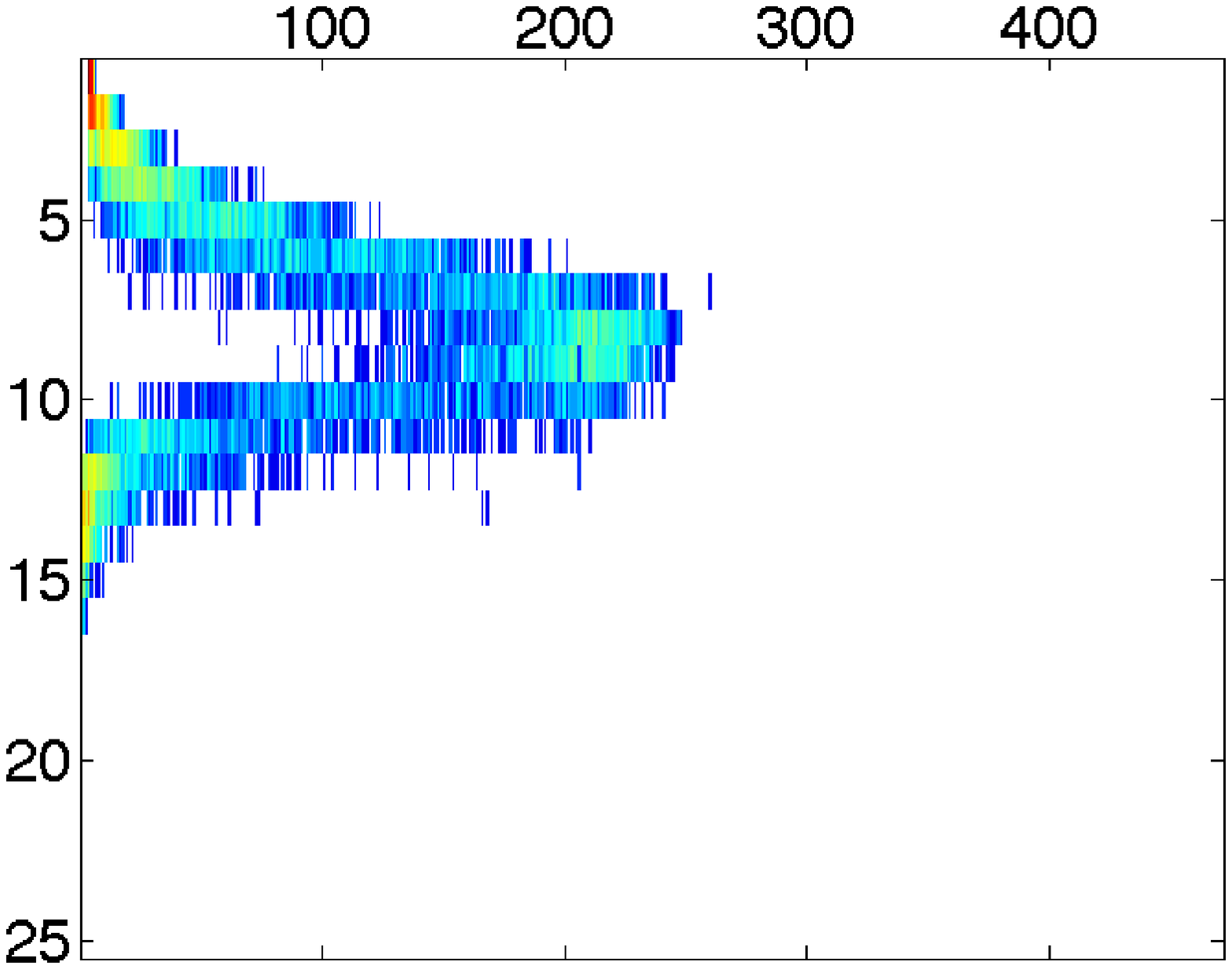}
	}
	\subfigure[][]{
	\label{subfig:NWS_1over6}
	    \includegraphics[trim=8 4 8 4,clip,width=0.22\textwidth]{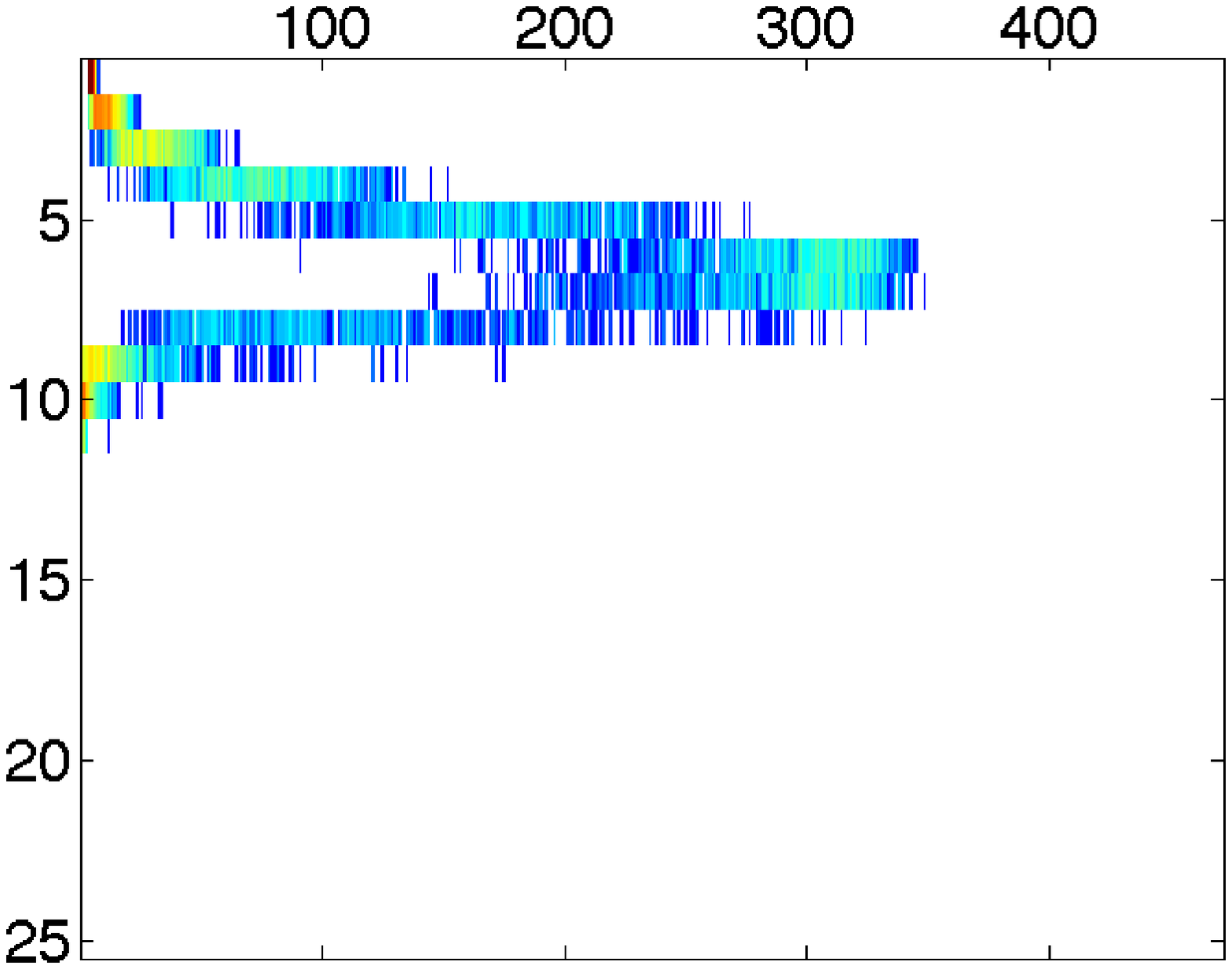}
	}
	\subfigure[][]{
	\label{subfig:NWS_1over3}
	    \includegraphics[trim=8 4 8 4,clip,width=0.22\textwidth]{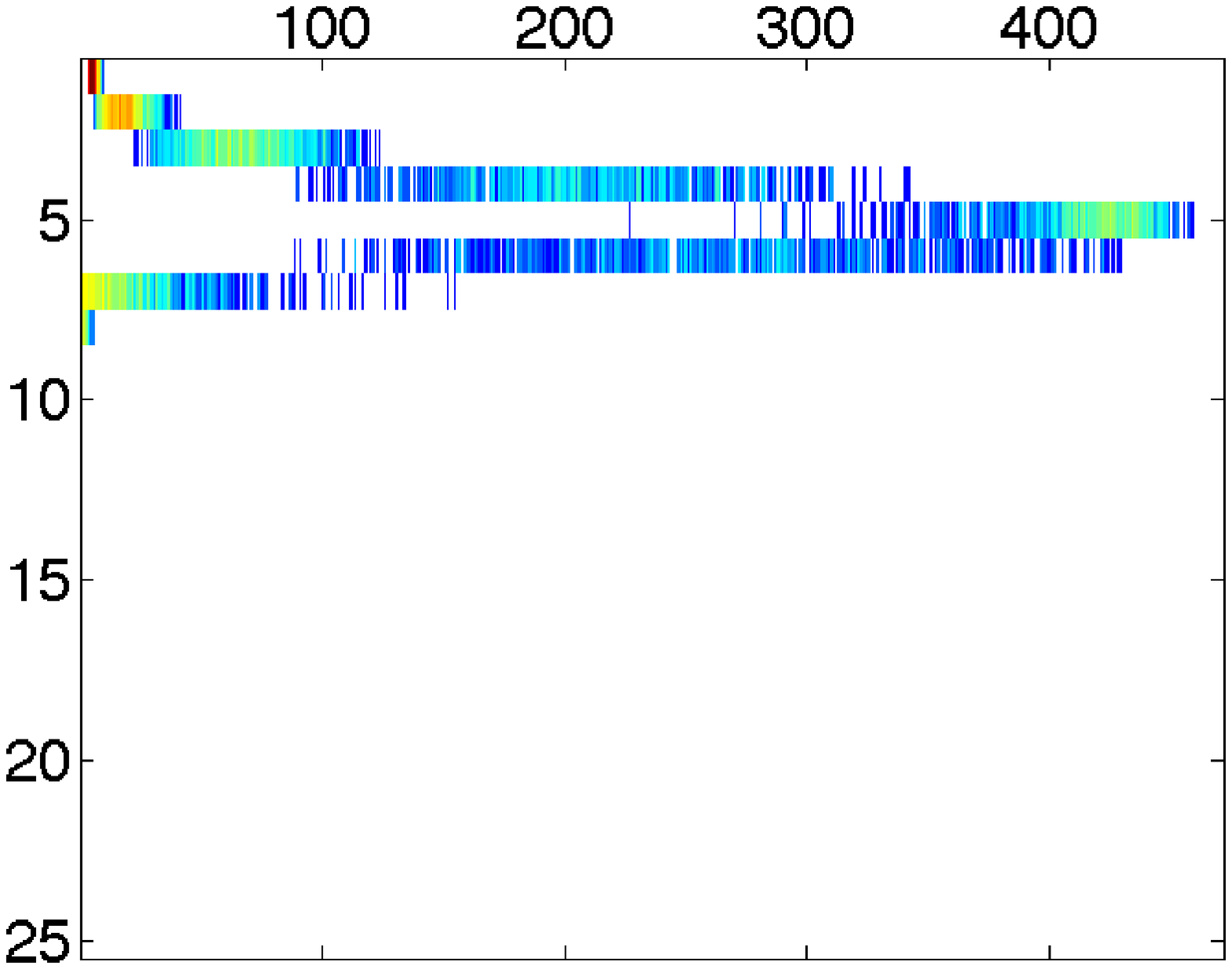}
	}
	\caption[]{\label{fig:NWS_smallWorldEmergence}(color online) The emergence of small world.  Shown are Newman-Watts-Strogatz graphs~\protect{\cite{newmanWattsStrogatz:PRE}} with $N=1000$; $k=4$; and $p=1/20, 1/10, 1/5$, and $2/5$; \subref{subfig:NWS_1over12}--\subref{subfig:NWS_1over3}, respectively.  }
\end{figure}

\begin{figure} 
\centering
	\subfigure[][]{
	\label{subfig:aveKR}
	    \includegraphics[trim=8 4 8 4,clip,width=0.22\textwidth,]{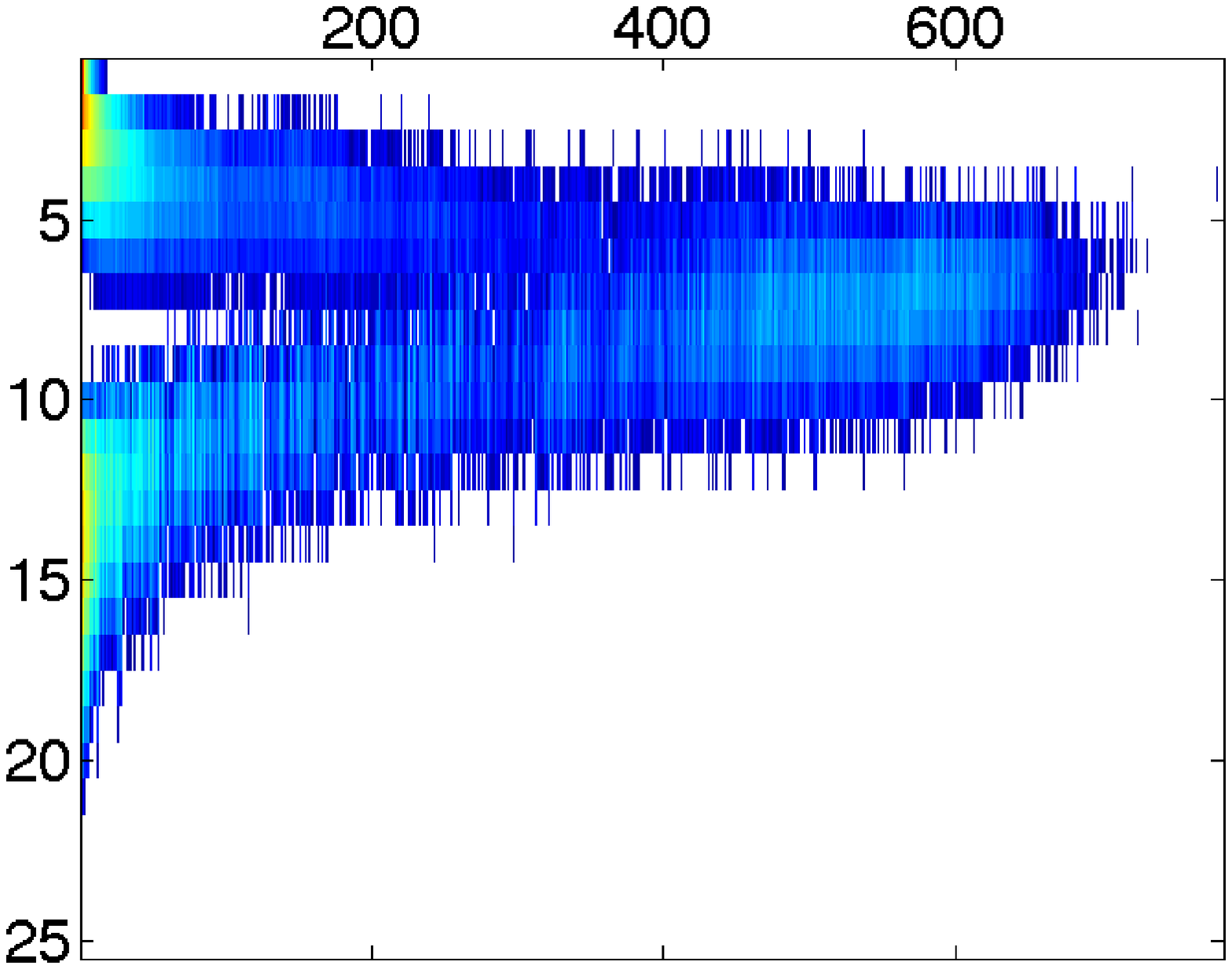}
	}
	\subfigure[][]{
	\label{subfig:aveBA}
	    \includegraphics[trim=8 4 8 4,clip,width=0.22\textwidth,]{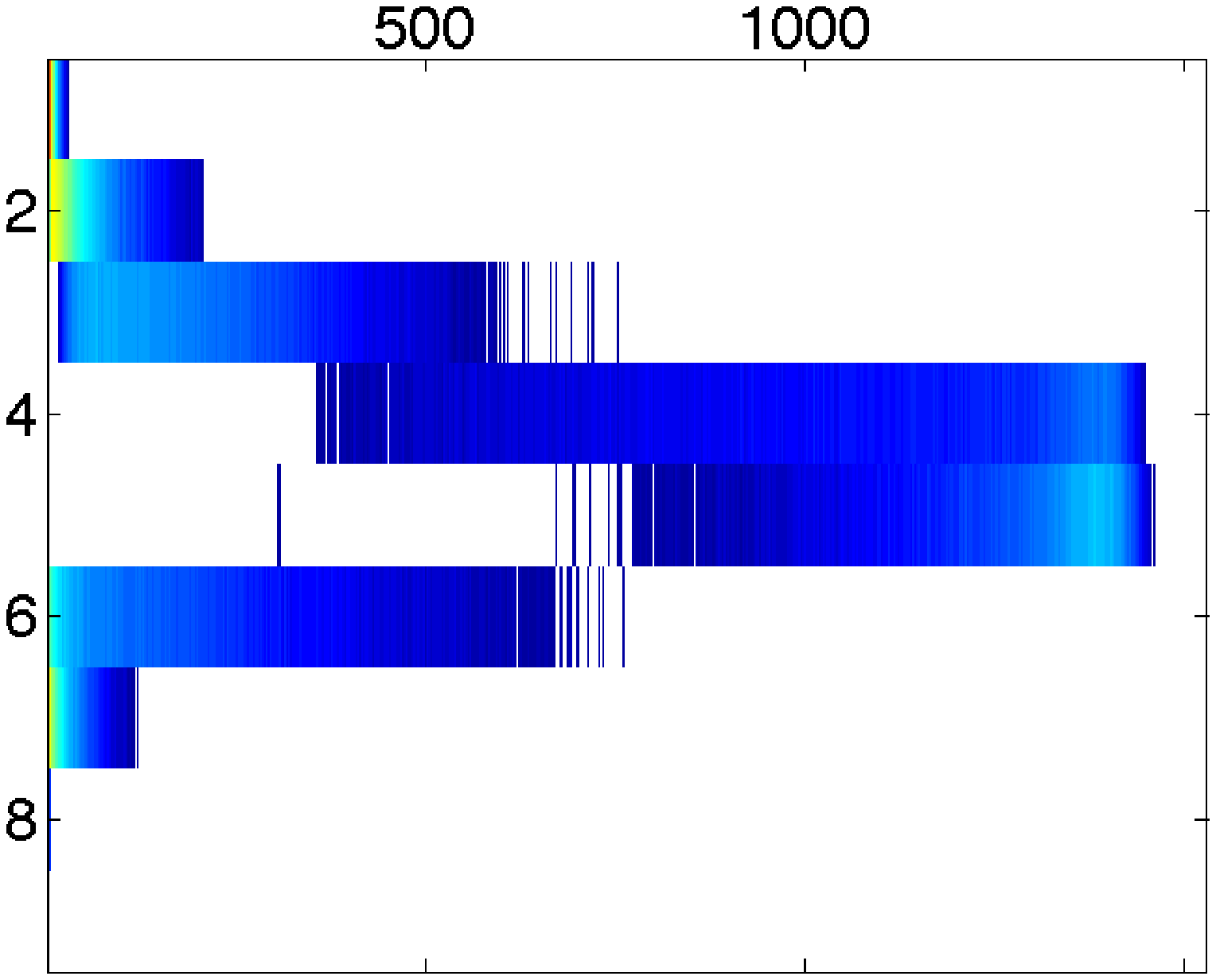}
	}
	\subfigure[][]{
	\label{subfig:aveMR}
	    \includegraphics[trim=8 4 8 4,clip,width=0.22\textwidth,]{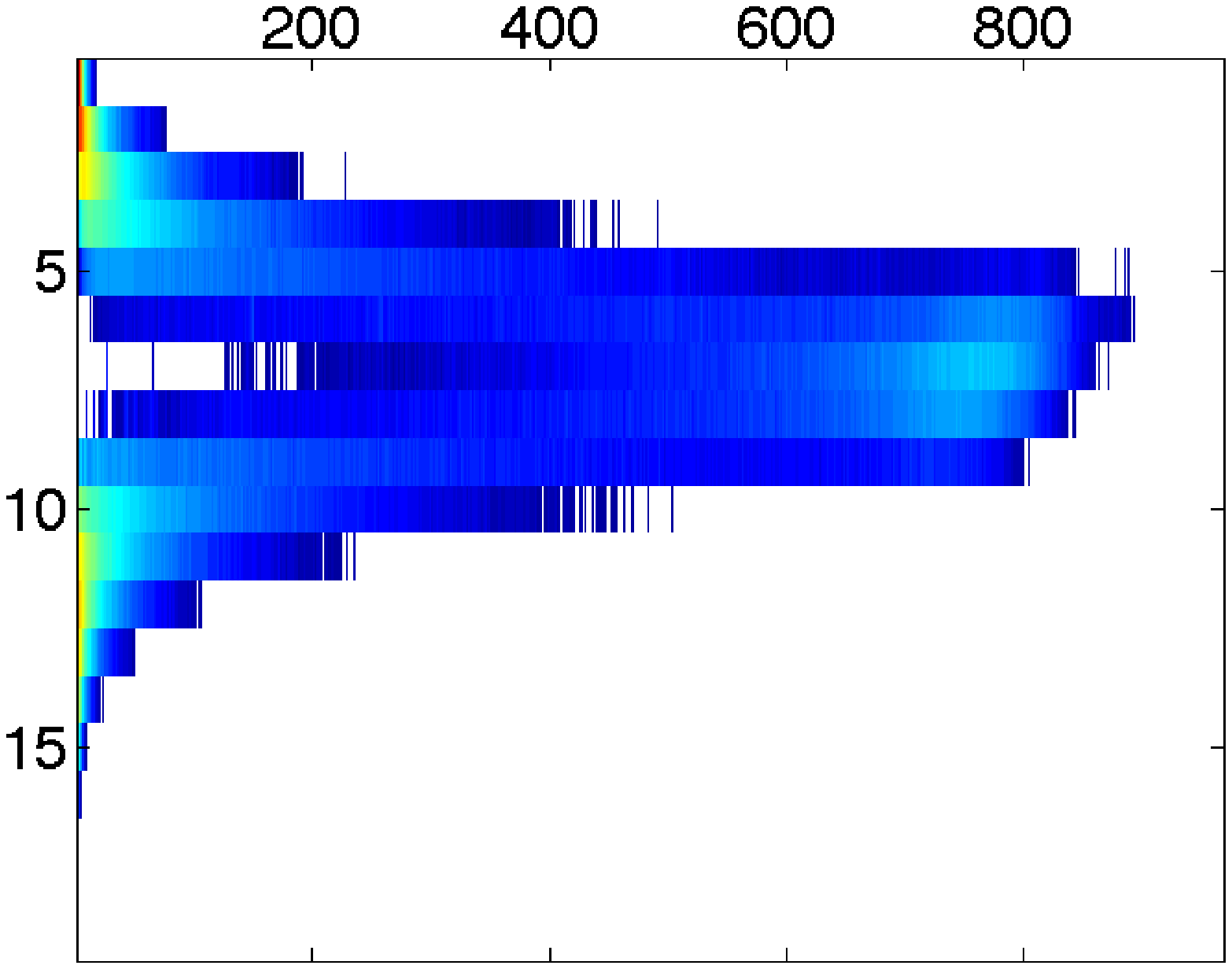}
	}
	\subfigure[][]{
	\label{subfig:UVflower}
	    \includegraphics[trim=8 4 8 4,clip,width=0.22\textwidth,]{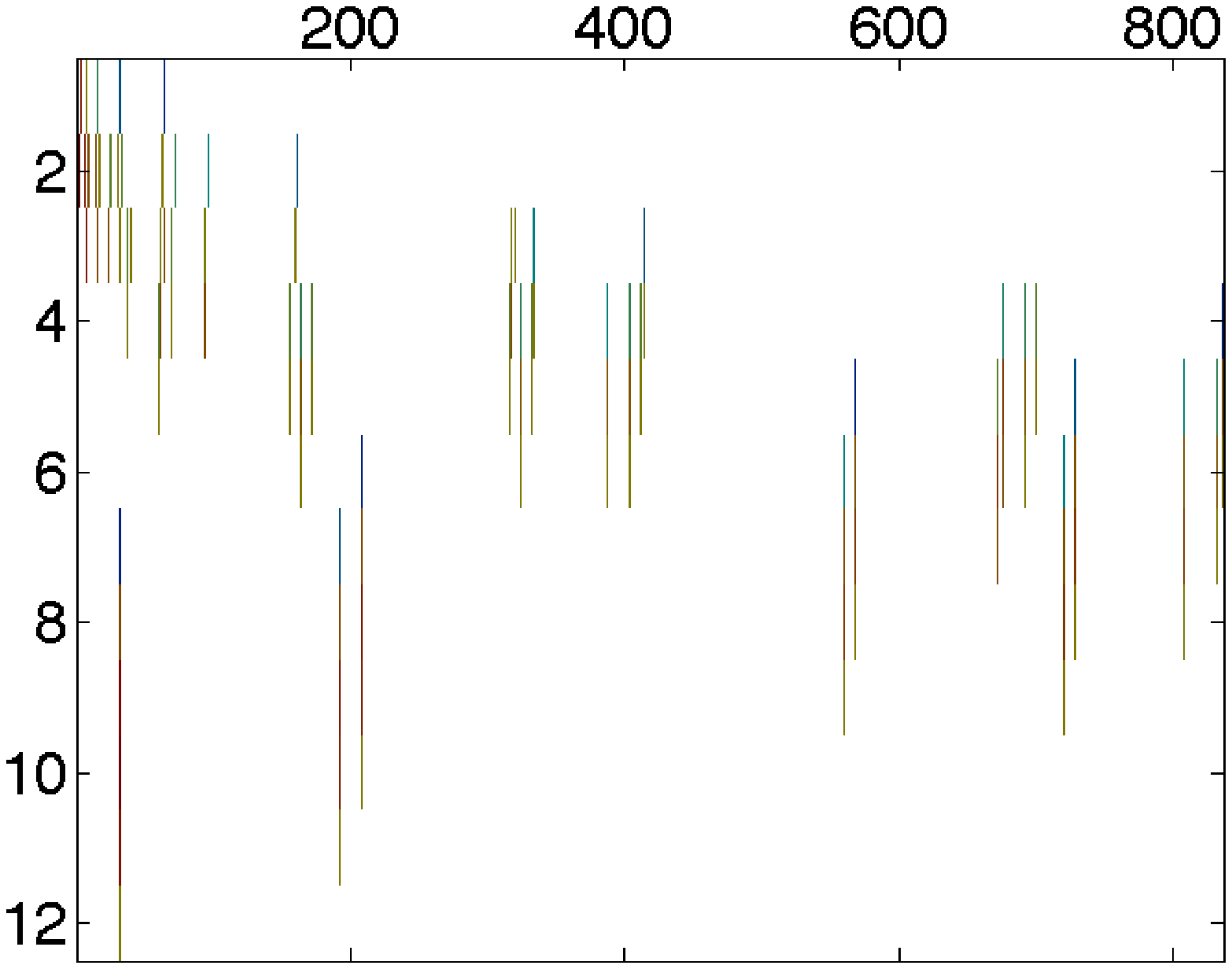}
	}
	\caption[]{\label{fig:ScaleFreeNets}(color online) Scale-Free models.  The average of 100 instances of the (undirected) Krapivsky-Redner ($r=1/2$)~\protect{\cite{krapivskyRedner}}; Barab\'{a}si-Albert (BA) ($m=2$)~\protect{\cite{barabasiBAmodel}}; and Molloy-Reed (MR) (drawn from $P(k) \sim k^{-3}$)~\protect{\cite{MR4}} networks; as well as the (1,3)-Flower at generation 6~\protect{\cite{rozenfeldFlowerNets}}; \subref{subfig:aveKR}--\subref{subfig:UVflower}, respectively. All have $N = 2732$, $\gamma \approx 3$, but $\left<k\right>$ varies.  Note that \subref{subfig:UVflower} has been darkened slightly for clarity.}
\end{figure}

\begin{figure}[t] 
\centering
	\subfigure[][]{
	\label{subfig:PeriodicSquareLattice}
	    \includegraphics[trim=8 4 8 4,clip,width=0.22\textwidth]{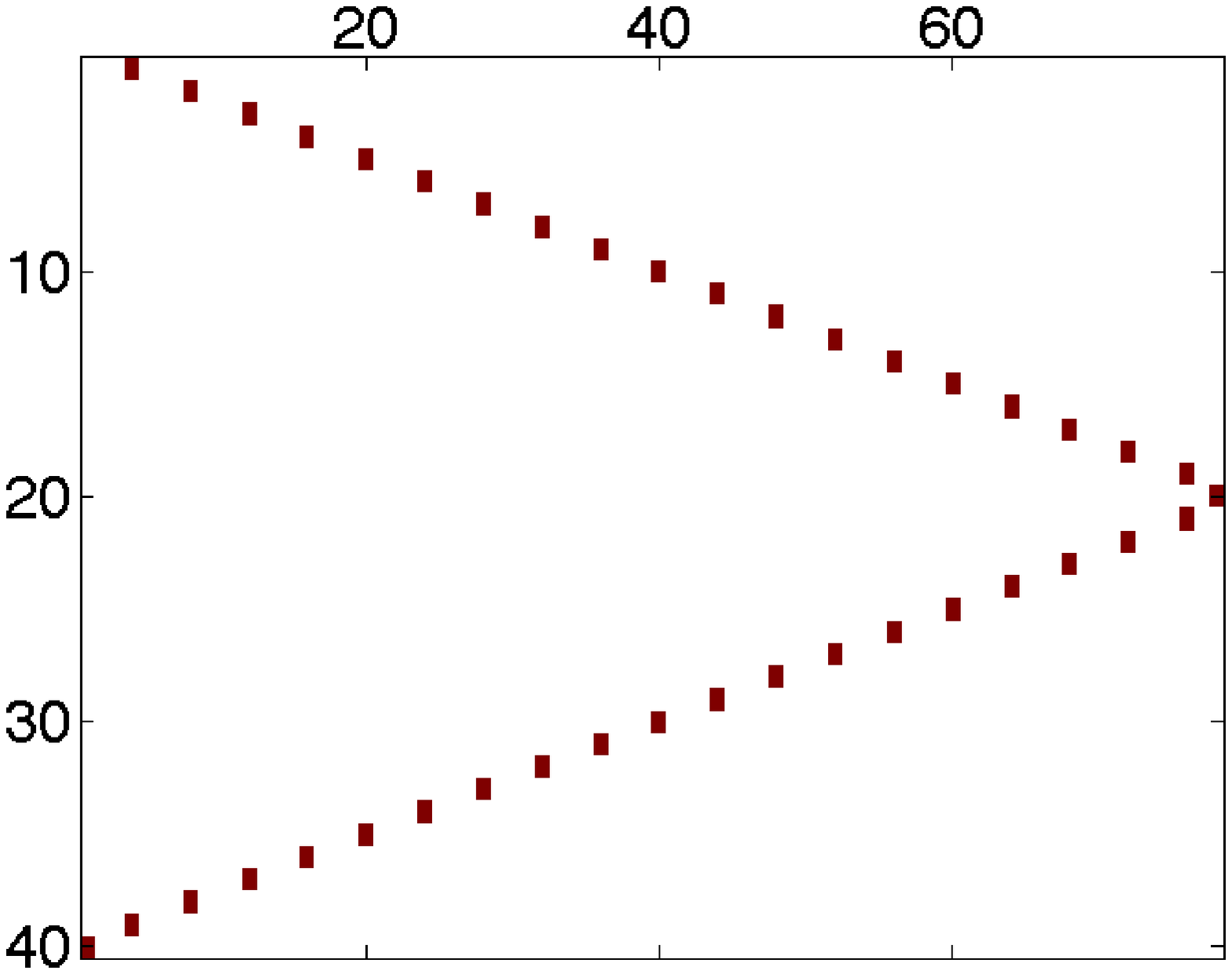}
	}
	\subfigure[][]{
	\label{subfig:nonPeriodicSquareLattice}
	    \includegraphics[trim=8 4 8 4,clip,width=0.22\textwidth]{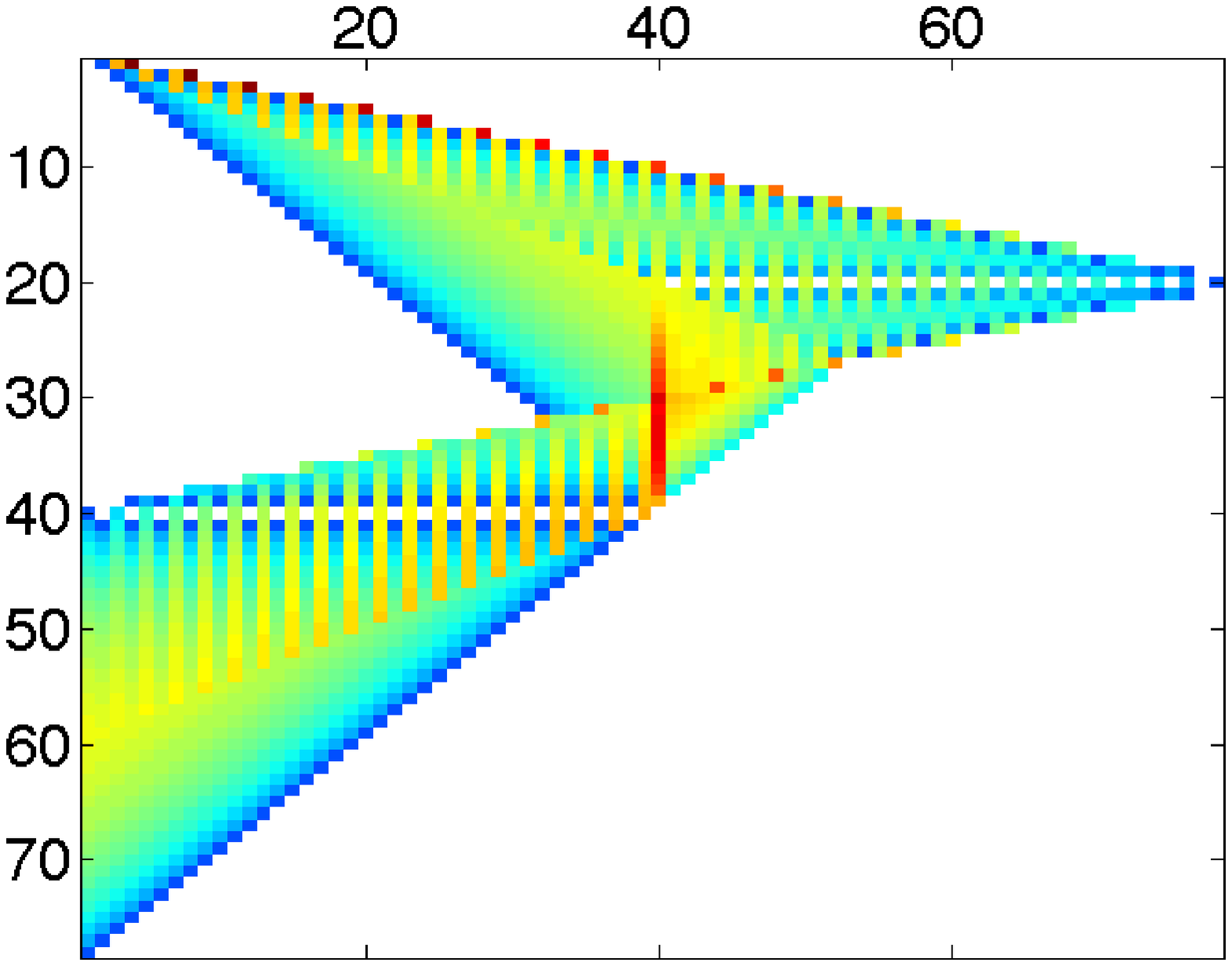}
	}
	\subfigure[][]{
	\label{subfig:skewedBC}
	    \includegraphics[trim=8 4 8 4,width=0.22\textwidth]{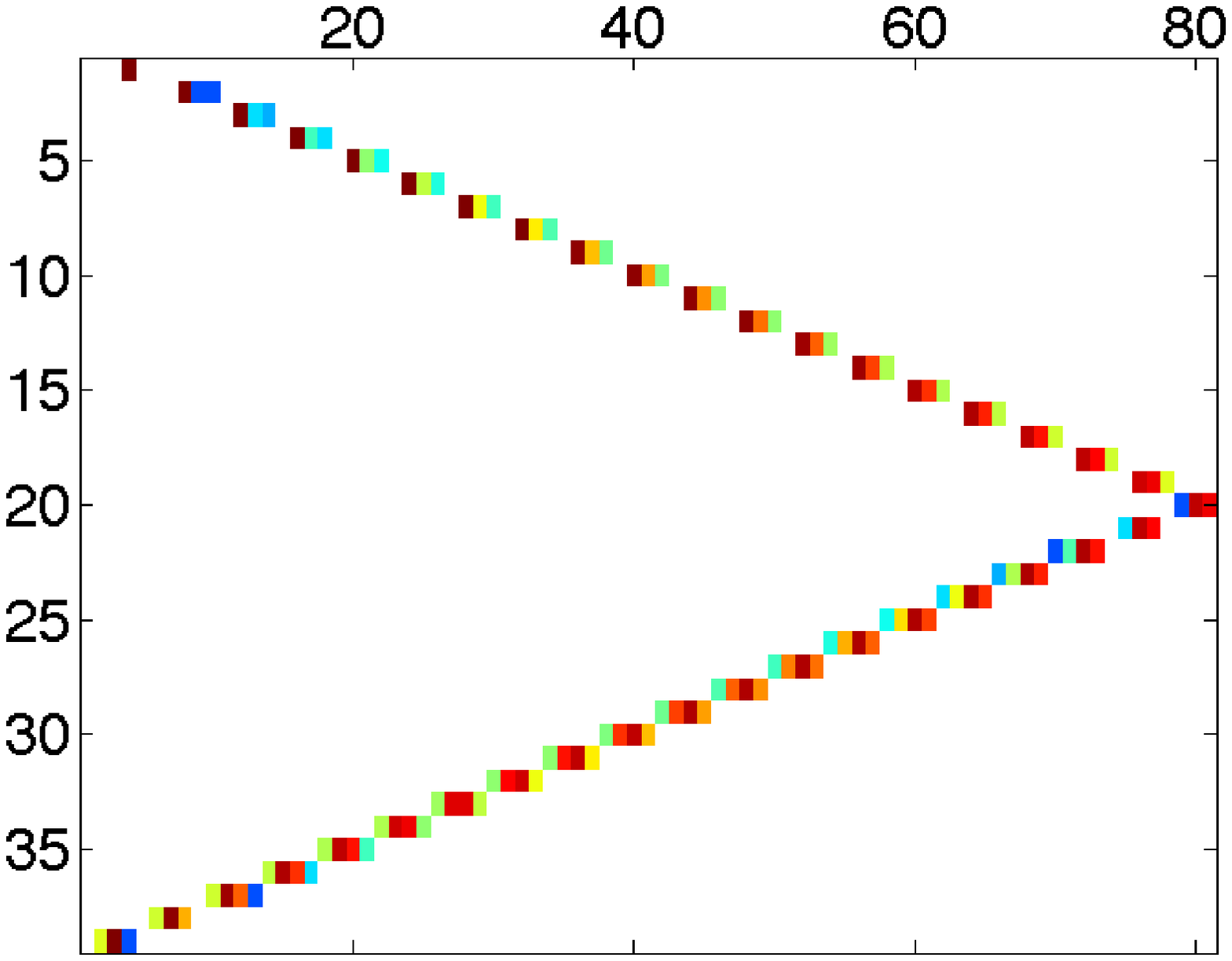}
	}
	\subfigure[][]{
	\label{subfig:5per_nodes}
	    \includegraphics[trim=8 4 8 4,clip,width=0.22\textwidth]{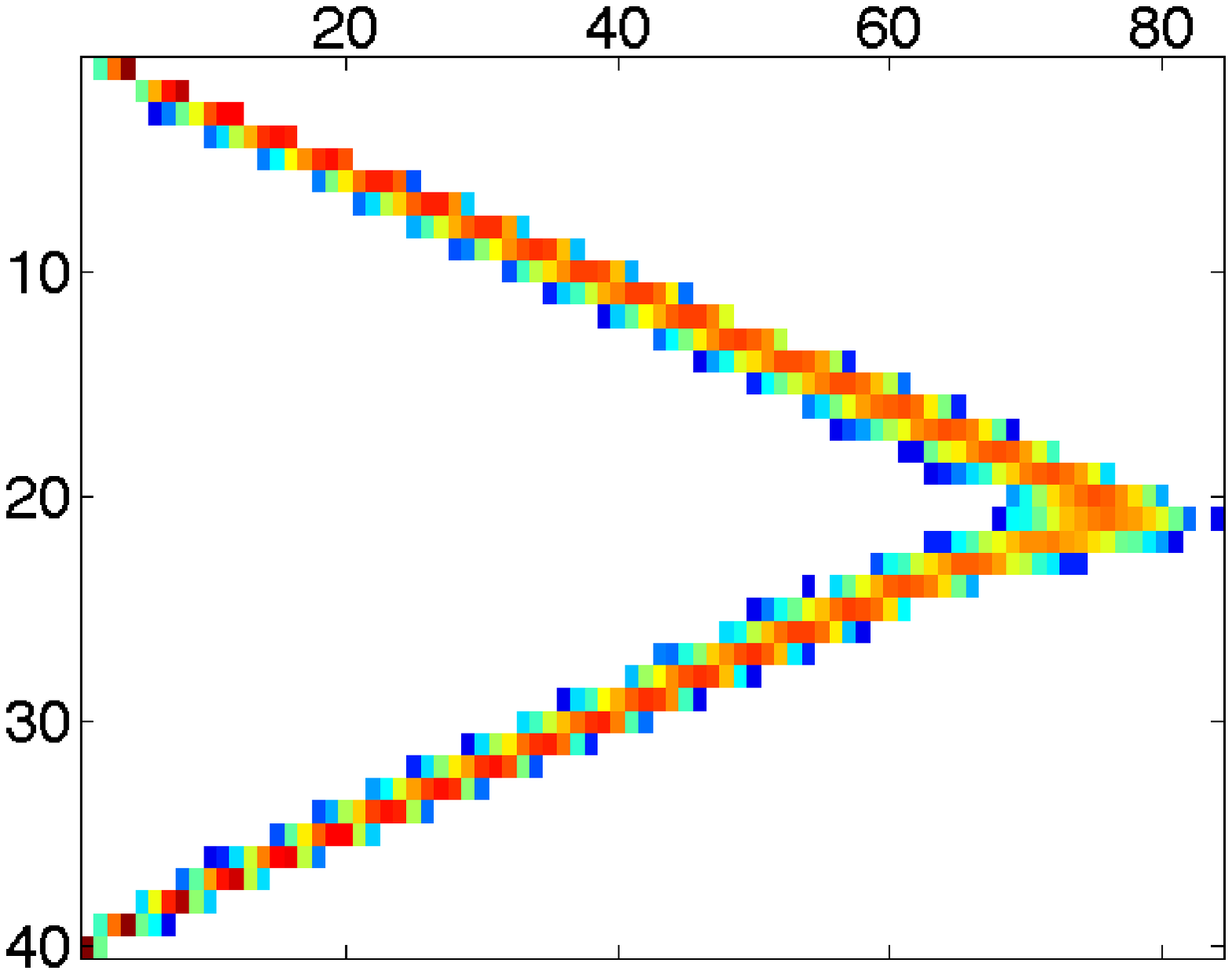}
	}
	\caption[]{\label{fig:variousLattices}(color online) Regular $40 \times 40$ lattices with defects. \subref{subfig:PeriodicSquareLattice} A periodic and \subref{subfig:nonPeriodicSquareLattice} non-periodic lattice; \subref{subfig:skewedBC} a lattice with skew-periodic boundaries; and~\subref{subfig:5per_nodes} a periodic lattice with a random 5 percent of all nodes missing.  Observe the strong linear slope, indicating the underlying two-dimensional lattice, as well as the narrowness of the distributions in \subref{subfig:PeriodicSquareLattice}, \subref{subfig:skewedBC}, and \subref{subfig:5per_nodes}, due to the regularity of the periodic lattice. Similarly, 1D lattices show a constant (vertical) line and 3D lattices exhibit quadratic growth.}
\end{figure}

\begin{figure} 
\centering
	\subfigure[][]{
	\label{subfig:pow_un}
	    \includegraphics[trim=8 4 8 4,clip,width=0.22\textwidth]{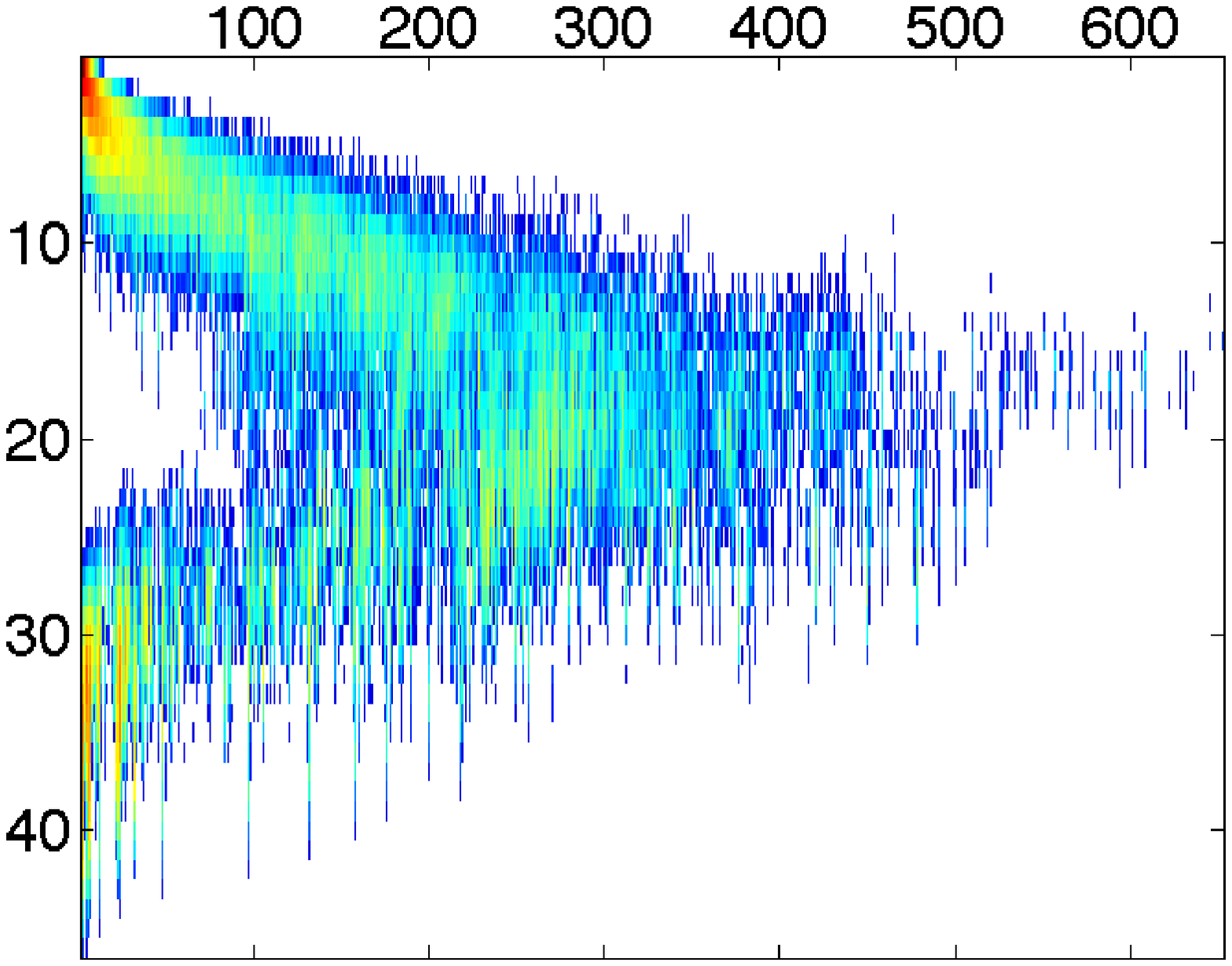}
	}
	\subfigure[][]{
	\label{subfig:CS}
	    	   \includegraphics[trim=8 4 8 4,clip,width=0.22\textwidth]{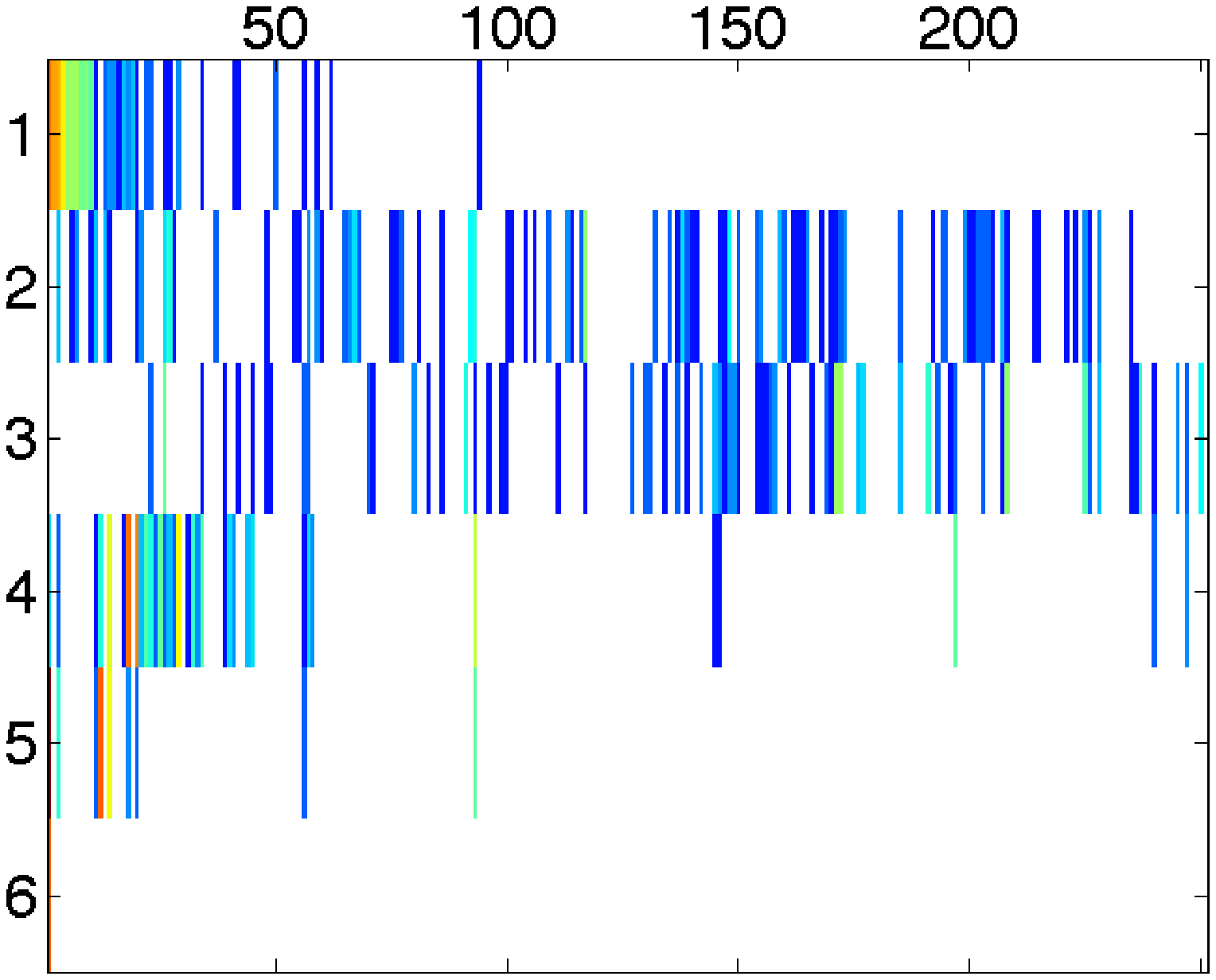}
	}
	\subfigure[][]{
	\label{subfig:RCd}
	    \includegraphics[trim=8 4 8 4,clip,width=0.22\textwidth]{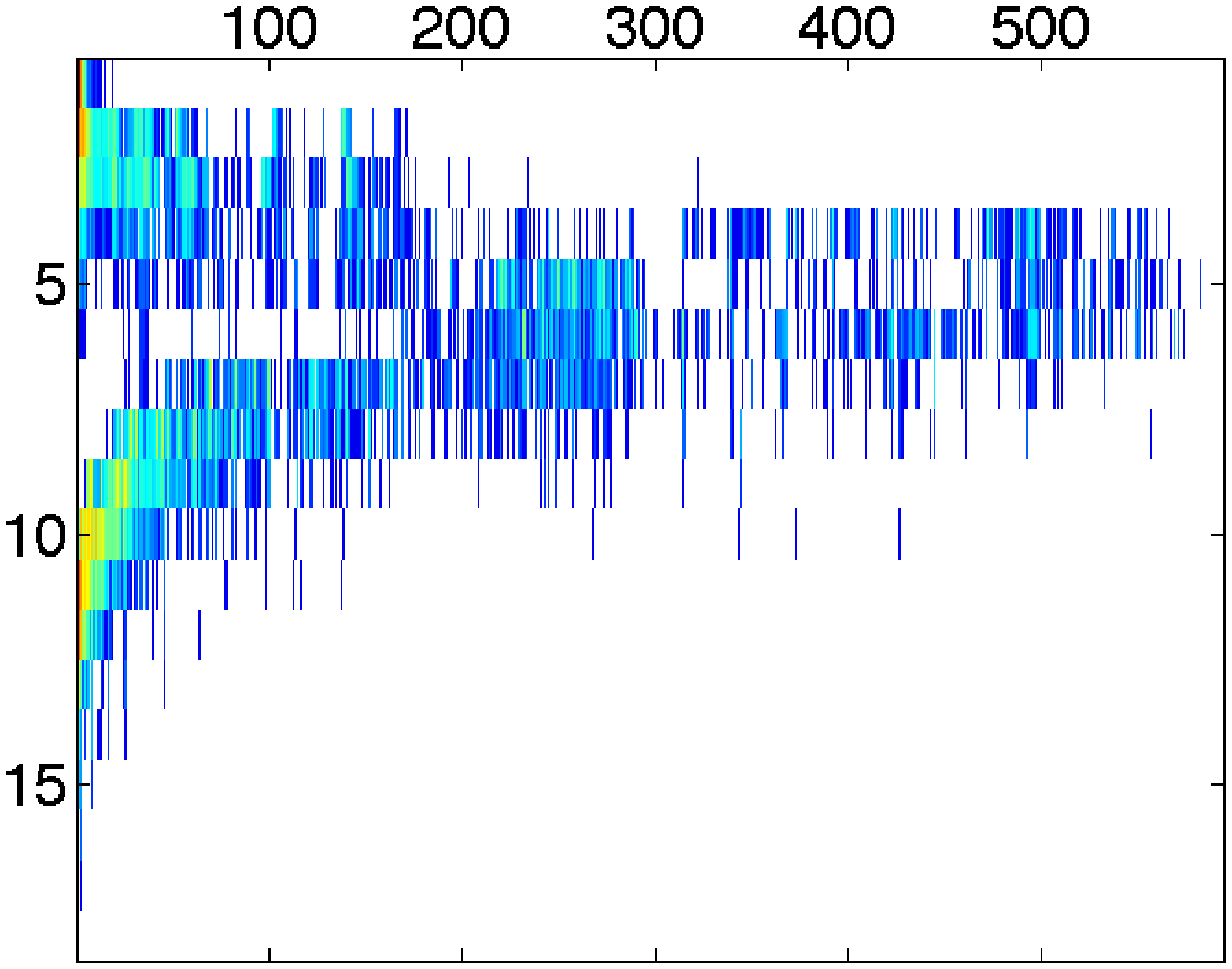}
	}
	\subfigure[][]{
	\label{subfig:HId}
	    \includegraphics[trim=8 4 8 4,clip,width=0.22\textwidth]{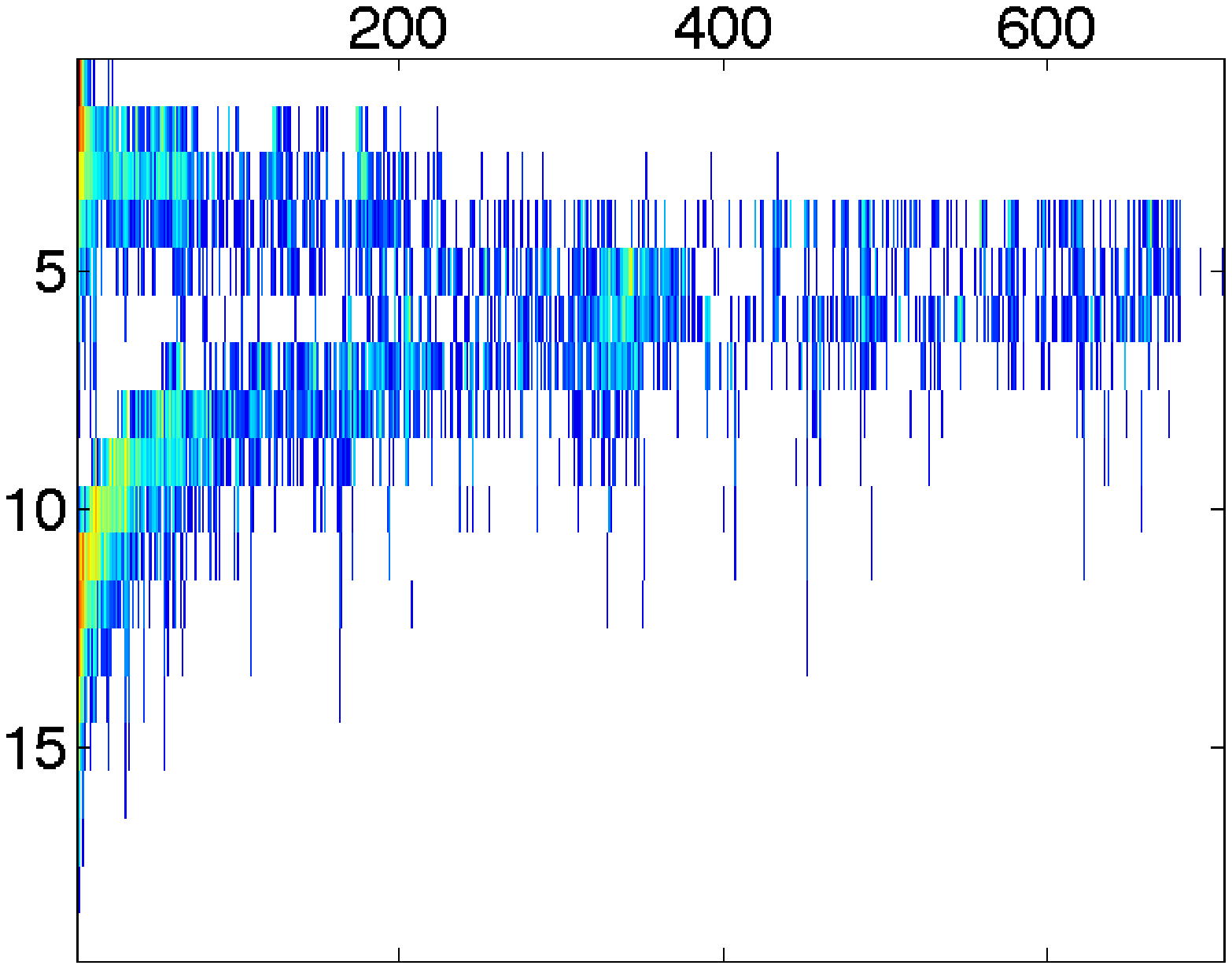}
	}
	\subfigure[][]{
	\label{subfig:CEd}
	    \includegraphics[trim=8 4 8 4,clip,width=0.22\textwidth]{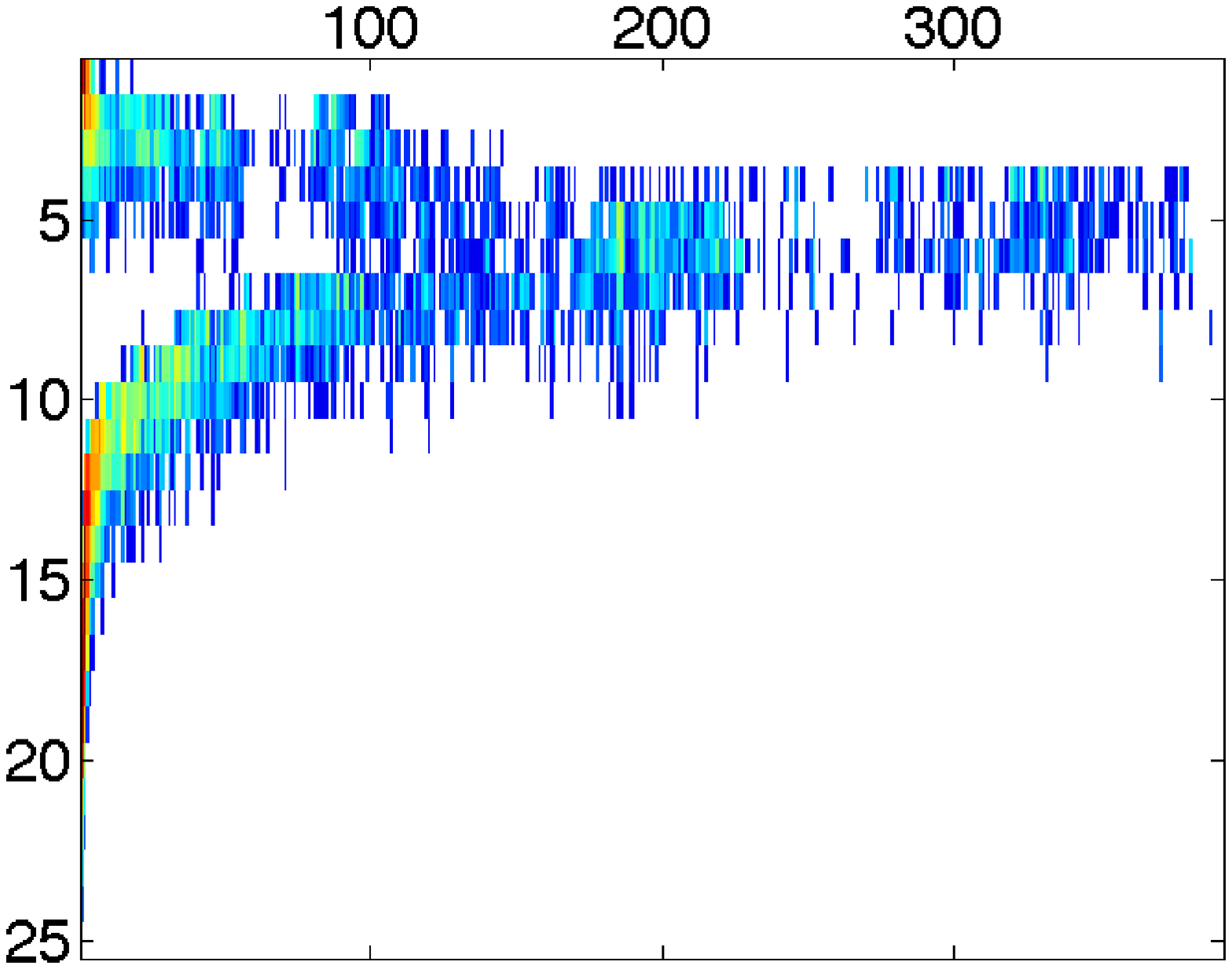}
	}
	\subfigure[][]{
	\label{subfig:MJd}
	    \includegraphics[trim=8 4 8 4,clip,width=0.22\textwidth]{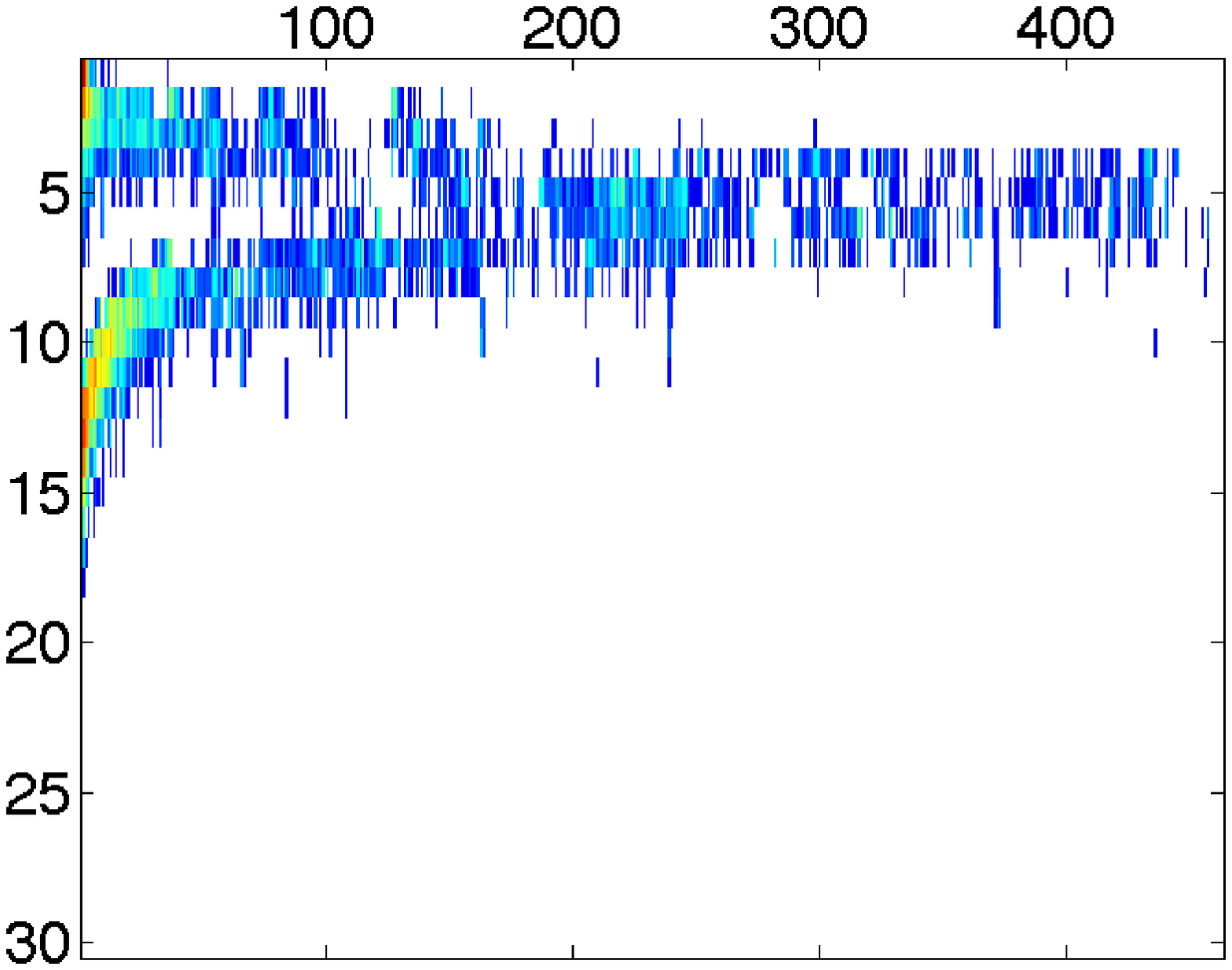}
	}
	\caption[]{\label{fig:BS_realWorld}(color online) Several real world networks.  \subref{subfig:pow_un} The western states power grid (unweighted)~\protect{\cite{wattsStrogatz:Nature:SmallWorld}}; \subref{subfig:CS} US airlines network~\protect{\cite{USAir97_1}}; and directed metabolic networks for \emph{H. influenzae}, \emph{R. capsulatus}, \emph{M. jannaschii}, and \emph{C. elegans}~\protect{\cite{BA:nature:metabolic}}, \subref{subfig:RCd}--\subref{subfig:MJd}, respectively.  The metabolic networks appear similar to one another yet unlike the power grid and airlines networks.}
\end{figure}

\begin{figure} 
\centering
\subfigure[][]{
\label{subfig:MGbeforeRewire}
    \includegraphics[trim=8 4 8 4,clip,width=0.22\textwidth]{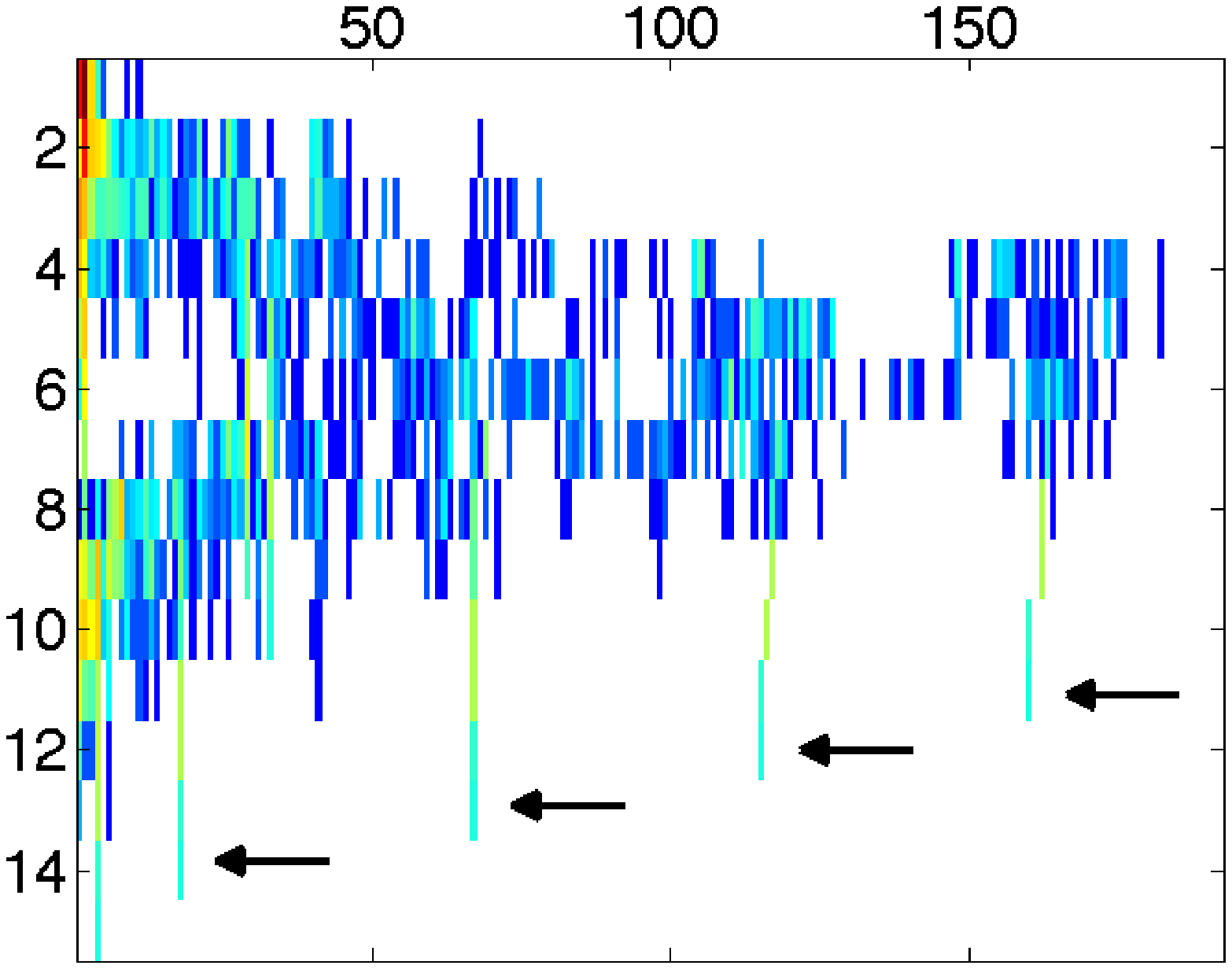}
}
\subfigure[][]{
\label{subfig:MGafterRewire}
    \includegraphics[trim=8 4 8 4,clip,width=0.22\textwidth]{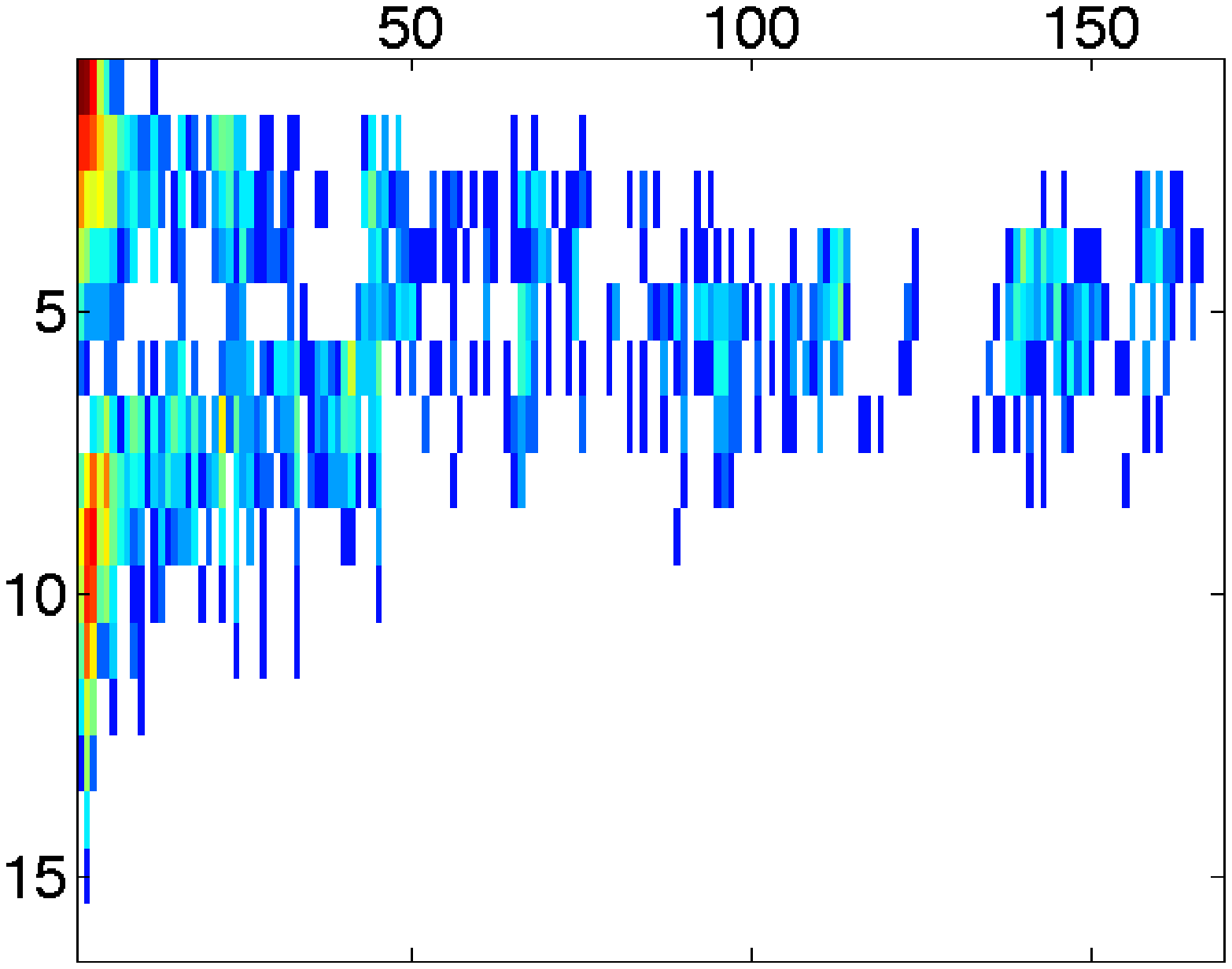}
}
\caption{(color online) \subref{subfig:MGbeforeRewire} The original metabolic network of \emph{M. genitalium}~\protect{\cite{BA:nature:metabolic}} with assortativity $A = -0.174216$ and \subref{subfig:MGafterRewire} with $A = 0.000757$ after permuting random edge pairs while preserving the degree distribution.  The fine-scale structure in the upper-most shells of \subref{subfig:MGbeforeRewire} is no longer present in \subref{subfig:MGafterRewire}.  }
\label{fig:assortRewire}
\end{figure}

We now introduce the  $B$-matrix.  Define the \emph{distance} between two nodes as the smallest number of links connecting them, found using Breadth-First Search (BFS)~\cite{NP}.  Thus, a node $v_i$ is surrounded by $\ell$-shells: the subsets of nodes at distance $\ell$ from $v_i$.
Let
\begin{eqnarray}
\label{Bmatrix}
B_{\ell,k} = \mbox{ number of nodes that have exactly $k$ } \\
 \mbox{ members in their respective $\ell$-shells.} \nonumber
\end{eqnarray}
Note that $B$ is independent of node labeling: all isomorphs of a graph have exactly the same $B$-matrix.  Enumerating the shell members of a specific node requires ${\cal O}(N)$ steps for a sparse graph~\cite{NP}, thus construction of the $B$-matrix requires ${\cal O}(N^2)$ steps.  Example $B$-matrices are shown in Figs.\ \ref{fig:introduceBmatrix}--\ref{fig:assortRewire} and discussed in the Results section.

It is easy to see, from~(\ref{Bmatrix}), that the degree distribution of a graph is encoded in the first row of its $B$-matrix,
\begin{equation}
	B_{1,k}=N P(k)\;,
\end{equation}
since the degree of a node equals the number of neighbors in its $\ell=1$ shell.   Generalizing this concept, we define the {\it degree of order} $\ell$ of a node as the number of members in its $\ell$-th shell.  Then, row $\ell$ of the $B$-matrix lists the graph's distribution of degrees of order $\ell$:
\begin{equation}
	B_{\ell,k}=N P_\ell(k)\;.
\end{equation}

Consider a {\it maximally random\/} network, constructed by the Molloy-Reed algorithm~\cite{MR4}. 
Its structure is fully determined by its (first-order) degree distribution, or by the first row of its $B$-matrix.  
For example, the second row is
\begin{equation}
\label{B2}
B_{2,k}=\sum_lB_{1,l}\!\!\!\!\!\!\!\!\!\!\!\!
\sum_{{j_1,j_2,\dots,j_l\atop j_1+j_2+\cdots+j_l=k+l}}\!\!\!\!\!\!\!\!\!(p_{j_1}+p_ {j_2}+\cdots+
p_{j_l})\,,
\end{equation}
where $p_m\equiv mB_{1,m}/\sum_n nB_{1,n}$.
Thus the $B$-matrix contains much additional information beyond the degree distribution, encoded in the difference between the actual $B_{2,k}$ and the expression~(\ref{B2}) (and similarly for higher rows).  

\section{Results} \label{sec:results}

The intuition one gains simply by \emph{looking} at these portraits is of great value~\cite{remark1}.  Classification and comparison are immediate (Figs.\ \ref{fig:ScaleFreeNets}, \ref{fig:BS_realWorld}).  Dimensionality and regularity are encoded in the overall slope and row variances (Fig.\ \ref{fig:variousLattices}), while small-world behavior is displayed in the ``aspect ratio'' (Fig.\ \ref{fig:NWS_smallWorldEmergence}).  Even correlation effects
are discernable in the fine scale structure of the higher rows (Fig.\ \ref{fig:assortRewire}).  Properties such as assortativity were previously impossible to visualize for even moderately sized networks.
 
Here is a list summarizing the contents and ``moral of the story'' for each panel, numbered by figure:

\begin{enumerate}

\item[2.] The algorithm is cheap enough to visualize very large matrices, as indicated by this example and its nearly 30000 columns.  This also shows that a large amount of information is present in the matrix, far beyond the degree distribution encoded in the first row.

\item[3.] A large random network's $B$-matrix looks like the average of an ensemble of such networks (of the same size) (panels a,b).  A phase transition such as percolation is immediately visible (c,d).

\item[4.] The transition to small world is visible in the changing `aspect ratio' of the portrait.  These portraits have all been padded to the same dimensions.

\item[5.] Scale-free networks with identical numbers of nodes and power law exponents can still give radically different portraits.  Thus the portrait can be used to infer a generating mechanism or scale-free model, by providing information beyond the degree distribution.

\item[6.] Lattice defects, dimensionality (since shells scale like the dimension - 1), and ``regularity'' are all visible in the portrait.  This is useful, since the change in edges between a periodic and non-periodic lattice is small, though very specific, and this leads to massive change in the corresponding portraits.  

\item[7.] Real world networks can give remarkably different portraits, but some classes of real-world networks can look similar (shown here with four metabolic networks in panels c--f).  The four metabolic networks look quite similar despite widely varying scales in both axes.  This suggests a simple scaling procedure;  stretch one or both axes until the portraits overlap.

\item[8.] Correlation effects may still be visible in the higher rows of the portrait.  Here is a highly disassortative metabolic network, note the vertical structures in the higher rows.  Rewiring or perturbing this network to raise the assortativity destroys these structures. 

\end{enumerate}

\section{Comparing portraits} \label{sec:comparing}

\begin{figure}[t] 
	\centering
	\begin{minipage}{0.75\columnwidth}
	\label{subfig:twoER}
		\includegraphics[width=\textwidth]{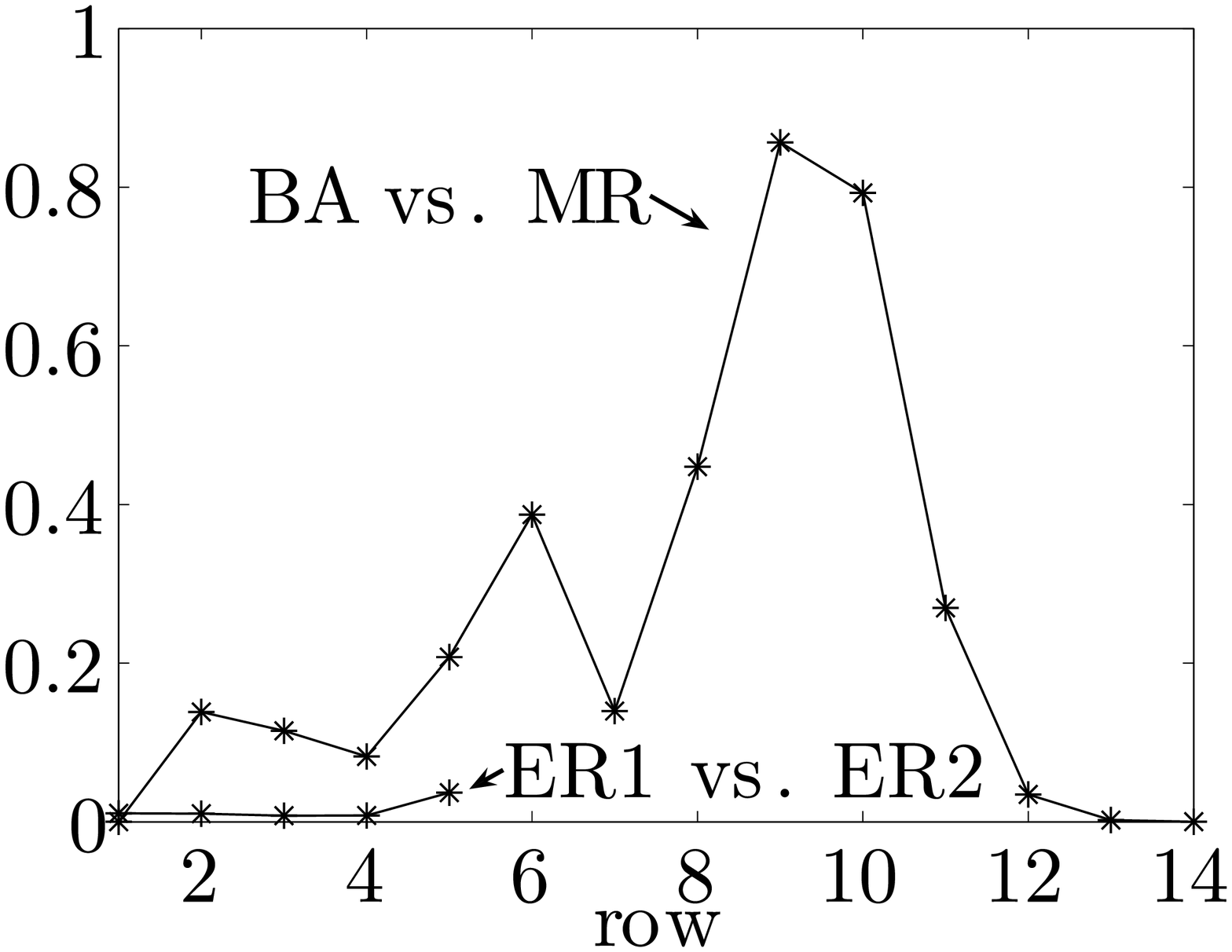}\\
	\end{minipage}
	\begin{minipage}{\columnwidth}
	\centering
		\begin{tabular}[b]{l|llll}
			 &  ER1    &   ER2  & BA   & MR  \\
			\hline
			ER1 &     0 & 0.012 & 0.654 & 0.620 \\
			ER2 & 0.012 &     0 & 0.654 & 0.619 \\
			BA  & 0.654 & 0.654 &     0 & 0.232 \\
			MR  & 0.620 & 0.619 & 0.232 &     0
		\end{tabular}
	\end{minipage}
\caption[]{\label{fig:RowWiseKSExamplePlots} (top) Row-wise statistic $K_{\ell}$:  two ER graphs with $N= 10^4$ and $p=0.002$; and a BA (diameter 10) versus an MR network ($P(k)\sim k^{-3}$, diameter 14), both with $N=5\times 10^4$. Both the BA and MR networks have the same degree distribution, so the first rows agree.  Differences in, e.g., assortativity, soon become evident.  (bottom) Table containing the values of $\Delta$, given by Eq.~(\ref{eqn:weightedMean}), for the four networks. This table shows that the two ER graphs are very close to each, while the MR and BA graphs are somewhat far apart from each other and very far from the ER graphs, as expected. } 
\end{figure}

The portraits are useful for showing an intuitive picture of a network, but they can also be used quantitatively.  A simple ``distance'' comparing networks $G$ and $G'$ may be defined, using their respective $B$-Matrices~\footnote{We assume that the networks are of comparable size.
Empirically, the $B$-matrices may be scaled and normalized: $\{\ell,k,B\} \mapsto \{\ell/L,k/K,B/N\}$, where $L$ and $K$ are the largest
shell number and largest degree (of any order), respectively.  When the number of rows is small, one may first replace the $B$-matrix by a suitably smoothed surface (applying a spline procedure), then proceed with scaling.}. Motivated by the Kolmogorov-Smirnov (KS) test~\cite{book:KStest}, we introduce the following statistic between corresponding pairs of rows $B_\ell$ and $B_\ell'$%
:
\begin{equation}
	K_\ell = \max_k \left| C^{}_{\ell,k} - C_{\ell,k}^{\prime}\right| , \label{eqn:K_l}
\end{equation}
where $C$ is the matrix of cumulative distributions of $B$, 
\begin{equation}
	C_{\ell,k} =\left( \sum_{k'\leq k}B_{\ell,k'} \right) / \sum_{k'} B_{\ell,k'} .
\end{equation}
The greater impact of lower shells on network properties (such as the average path length~\cite{newmanGeneratingFunctions,dorogovtsevShellNuclPhysB}) can be considered by assigning weights 
$\alpha_{\ell}$, based on shell ``mass,'' for instance:
\begin{equation}
	\alpha_\ell =\sum_k B_{\ell,k} + \sum_k B_{\ell,k}'.
	\label{eqn:possibleWeights}
\end{equation}
Finally, we choose a scalar distance $\Delta$, generated by
\begin{equation}
	\Delta(G,G') \equiv \Delta(B,B') = \left(\sum_\ell \alpha_\ell K_\ell\right) / \sum_\ell \alpha_\ell .
	\label{eqn:weightedMean}
\end{equation}  
See Fig.\ \ref{fig:RowWiseKSExamplePlots} for some concrete examples.

We apply this distance metric to four networks, summarized in Fig. 9.  Two  \ER (ER) networks, with equal $N$ and $p$, and a \BA (BA) versus a \MR (MR) network built from the BA degree distribution.  The plot indicates the value of the test statistic, Eq.~\ref{eqn:K_l}, while the table indicates the values of $\Delta$, from Eq.~\ref{eqn:weightedMean}.  The plot shows that the two ER networks agree very well with each other, while the BA and MR networks agree at first, but differences appear in higher rows (since BA has correlation effects missing in MR).  The table values all agree with expectations: the ER graphs are very close to each other, the BA and MR graphs are farther apart from each other, and both BA and MR are very far from the ER networks.

Mathematically, it remains an open question if $\Delta$ is a metric or semi-metric (pseudometric).  It is obvious from Eqs.~\ref{eqn:K_l} and \ref{eqn:possibleWeights} that $\Delta(x,y) \geq 0$ and $\Delta(x,y) = \Delta(y,x)$ . Furthermore, the numbers in Fig.~\ref{fig:RowWiseKSExamplePlots} satisfy the triangle inequality, but does this hold generally?  The final issue at hand concerns indiscernibility, $\Delta(x,y) = 0 \iff x=y$.  Discernibility in $\Delta(B,B')$ appears to hold, but there exist two non-isomorphic graphs, the dodecahedral and Desargues graphs, which have identical $B$'s, disproving discernibility in $\Delta(G,G')$, if only because their $B$-matrices are indiscernible~\footnote{These graphs are both discussed in the final section.}. 

\section{Conclusions and future work}
To summarize, $B$-matrices offer us an unambiguous way to visualize and discriminate between 
various complex networks.  With little practice one can readily pick the patterns that distinguish 
one case from another: for example, the metabolic networks (Fig.~\ref{fig:BS_realWorld}) have a distinctly similar appearance, with a prominent ``knot'' near the center of the portraits.  Even small changes in structure induce visible changes in the $B$-matrix (Figs.~\ref{fig:variousERgraphs}c,d, and \ref{fig:variousLattices}d); the largest changes being induced by the removal or addition of links of highest {\it betweenness centrality\/}~\cite{betweenness}.

We have also introduced a distance, associated with the $B$-matrix, that quantifies the differences between complex networks.  The distance between networks belonging to the same ensemble is small (Figs.\ \ref{fig:variousERgraphs}a,b, and \ref{fig:RowWiseKSExamplePlots}), but it grows larger for networks in different ensembles (Fig.\ \ref{fig:RowWiseKSExamplePlots}).

Several generalizations come to mind.
Eq.~(\ref{Bmatrix}) encompasses directed graphs and may be 
extended to weighted graphs: shells are defined by a set of weights $W = \{w_1, w_2, \ldots, w_d \}$ and could be found by Dijkstra's algorithm~\cite{DijkstrasAlgorithm}.  One may also generalize $B$ to \emph{edges} by defining the distance from a node $v_i$ to an edge $(v_j, v_k)$ as the mean of distances $d(v_i,v_j)$ and $d(v_i,v_k)$~\footnote{$B_{\ell,k}$ is now the \# of nodes with $k$ \emph{edges} at distance $\ell$.}.  This ``edges matrix'' has half-integer rows with row $1/2$ encoding the degree distribution, $B_{1/2,k}=N P(k)$, and so forth.  

Among the most promising applications of $B$-matrices, besides identification and comparisons, is the question of the information content of complex graphs.  The portraits can be compressed by applying conventional algorithms.  The size of the compressed files could serve as a measure of
information content  (the difference in entropy of stochastic scale-free networks, vs.~that of the highly ordered flower, in Fig.~\ref{fig:ScaleFreeNets}, is apparent even visually).  

Another interesting problem would be to use the ``smoothness'' of the matrices to create some quantitative measure of regularity, perhaps based on the variances of each row.  This could also provide a useful measure of information content as well as symmetry and perhaps other characteristics.  In several instances, we have discussed the ``slope'' of the matrix without giving specifics.  While it's easy to identify dimensionality from the lattices of Fig.~\ref{fig:variousLattices}, other networks are of higher dimension with broader row distributions and it's more difficult to pick out the slope visually.  A specific fitting procedure or other technique may be useful.

Given the degree probability distribution (the first row of the $B$-matrix) there exist algorithms to construct complex networks that satisfy that degree distribution~\cite{MR4}.  Perhaps the most important open question is the inverse of obtaining the B-matrices: Given a $B$-matrix, find a procedure to construct random complex nets belonging to the ensemble represented by it. This is related to the question of satisÞability Ñ constructing a random net that satisÞes just the $P_{1}(k)$ degree distribution is already non-trivial~\cite{MR42}, and this complicates as higher-order $P_{j}$'s are added in. There exist already examples of procedures for obtaining maximally random nets with more than the $P_{1} (k)$ constraint, for example in \cite{weberPorto} it is shown how to satisfy both the degree distribution and arbitrary degree-degree correlations.

Regarding the famous Graph Isomorphism problem, consider the non-isomorphic dodecahedral and Desargues graphs; both are cubic distance-regular with 20 nodes~\cite{book:distanceRegular} and both have identical $B$-matrices~\footnote{Distance-regular graphs will have exactly one nonzero element per row in $B$; in principle, this may be exploited to \emph{search} for undiscovered distance-regular graphs by rewiring edges along some scheme to minimize the number of nonzero elements per row.  This would likely be cost-prohibitive in practice.}, so $B$ does not uniquely encode a network.  In practice, the probability of two large, non-isomorphic graphs chosen from a large ensemble having identical $B$-matrices appears to be vanishingly small, since the slightest difference will propagate throughout $B$.  The dodecahedral and Desargues graphs are very similar in appearance, and the specific relationship between their edge sets that allows for identical $B$'s is unlikely to arise at random.  We propose that $B$ is a ``very good'' answer to graph isomorphism.  It is also worth noting that the Desargues and dodecahedral graphs have different edge matrices: we conjecture that graphs are uniquely identified with both matrices.  The true power of $B$ as a measure of graph isomorphism remains an open question and warrants further study.

Finally, it is worth noting that the construction of $B$ requires an $\mathcal{O}(N^{2})$ algorithm, which may preclude its use for extremely large networks.  However, this algorithm is easily parallelized by spreading the starting nodes over multiple machines.

\acknowledgments
We thank H. Rozenfeld for discussions.  We thank the NSF for support from a National Science Foundation Graduate Research Fellowship (JPB) and awards no.~DMS-0404778 (EB) and  PHY0555312 (DbA).

\end{document}